\documentclass[longauth]{aa}

\usepackage{graphicx}
\usepackage{color}
\usepackage[hyperfootnotes=false]{hyperref}
\hypersetup{colorlinks=True,linkcolor=blue,citecolor=blue, urlcolor=blue}
\usepackage{tablefootnote}
\usepackage{txfonts}

\newcommand{\cii}{[C\,{\sc ii}]\ }
\newcommand{\nii}{[N\,{\sc ii}]\,205$\mu$m\ }
\newcommand{\oiii}{[O\,{\sc iii}]\,88$\mu$m\ }

\begin{document} 

    \title{The ALMA-CRISTAL Survey} 

    \subtitle{Spatial extent of \cii line emission in star-forming galaxies at $z=4-6$}

    \author{
    Ryota Ikeda \inst{\ref{sokendai},\ref{naoj}}, 
    Ken-ichi Tadaki \inst{\ref{hgu}},
    Ikki Mitsuhashi \inst{\ref{naoj},\ref{uot}},
    Manuel Aravena \inst{\ref{udp}},
    Ilse De Looze \inst{\ref{ghent}}, \\
    Natascha M. F\"{o}rster Schreiber \inst{\ref{mpe}},
    Jorge Gonz\'{a}lez-L\'{o}pez \inst{\ref{udp},\ref{puc},\ref{lascampanas}},
    Rodrigo Herrera-Camus \inst{\ref{udc}},
    Justin Spilker \inst{\ref{tamu}}, \\
    Loreto Barcos-Mu\~{n}oz \inst{\ref{nrao},\ref{uov}},
    Rebecca A. A. Bowler\inst{\ref{uoman}},
    Gabriela Calistro\,\,Rivera\inst{\ref{dlr}},
    Elisabete da Cunha \inst{\ref{icrar},\ref{astro3d}}, \\
    Rebecca Davies \inst{\ref{astro3d},\ref{CfAS}}, 
    Tanio D\'{i}az-Santos \inst{\ref{forth},\ref{euc}}, 
    Andrea Ferrara \inst{\ref{sns}},
    Meghana Killi \inst{\ref{udp}},
    Lilian L. Lee \inst{\ref{mpe}},
    Juno Li \inst{\ref{icrar}}, \\
    Dieter Lutz \inst{\ref{mpe}},
    Ana Posses \inst{\ref{udp},\ref{tamu}},
    Renske Smit \inst{\ref{ljmu}},
    Manuel Solimano \inst{\ref{udp}}, 
    Kseniia Telikova \inst{\ref{udp}},
    Hannah \"{U}bler \inst{\ref{kic},\ref{cambridge}}, \\
    Sylvain Veilleux \inst{\ref{uom},\ref{jssi}},
    and Vicente Villanueva \inst{\ref{udc}}
    }
    \institute{
    Department of Astronomy, School of Science, SOKENDAI (The Graduate University for Advanced Studies), 2-21-1 Osawa, Mitaka, Tokyo 181-8588, Japan \email{ryota195ikeda@gmail.com} \label{sokendai} 
    \and National Astronomical Observatory of Japan, 2-21-1 Osawa, Mitaka, Tokyo 181-8588, Japan \label{naoj}
    \and Faculty of Engineering, Hokkai-Gakuen University, Toyohira-ku, Sapporo 062-8605, Japan \label{hgu}
    \and Department of Astronomy, The University of Tokyo, 7-3-1 Hongo, Bunkyo, Tokyo 113-0033, Japan \label{uot} 
    \and Instituto de Estudios Astrof\'{i}sicos, Facultad de Ingenier\'{i}a y Ciencias, Universidad Diego Portales, Avenida Ejercito Libertador 441, Santiago, Chile \label{udp}
    \and Sterrenkundig Observatorium, Ghent University, Krijgslaan 281 - S9, B-9000 Ghent, Belgium \label{ghent}
    \and Max-Planck-Institut f\"{u}r extraterrestrische Physik, Gie\ss enbachstra\ss e 1, 85748 Garching \label{mpe}
    \and Instituto de Astrof\'{i}sica, Facultad de F\'{i}sica, Pontificia Universidad Cat\'{o}lica de Chile, Santiago 7820436, Chile \label{puc}
    \and Las Campanas Observatory, Carnegie Institution of Washington, Ra\'{u}l Bitr\'{a}n 1200, La Serena, Chile \label{lascampanas}
    \and Departamento de Astronom\'{i}a, Universidad de Concepci\'{o}n, Barrio Universitario, Concepci\'{o}n, Chile \label{udc}
    \and Department of Physics and Astronomy and George P. and Cynthia Woods Mitchell Institute for Fundamental Physics and Astronomy, Texas A\&M University, 4242 TAMU, College Station, TX 77843-4242, US \label{tamu}
    \and National Radio Astronomy Observatory, 520 Edgemont Road, Charlottesville, VA 22903, USA \label{nrao}
    \and Department of Astronomy, University of Virginia, 530 McCormick Road, Charlottesville, VA 22903, USA \label{uov}
    \and Jodrell Bank Centre for Astrophysics, Department of Physics and Astronomy, School of Natural Sciences, The University of Manchester, Manchester, M13 9PL, UK \label{uoman}
    \and German Aerospace Center (DLR), Institute of Communications and Navigation, Wessling, Germany \label{dlr}
    \and International Centre for Radio Astronomy Research (ICRAR), The University of Western Australia, M468, 35 Stirling Highway, Crawley, WA 6009, Australia \label{icrar}
    \and ARC Center of Excellence for All Sky Astrophysics in 3 Dimensions (ASTRO 3D), Australia \label{astro3d}
    \and Centre for Astrophysics and Supercomputing, Swinburne University of Technology, Hawthorn, Victoria 3122, Australia \label{CfAS}
    \and Institute of Astrophysics, Foundation for Research and Technology-Hellas (FORTH), Heraklion 70013, Greece \label{forth}
    \and School of Sciences, European University Cyprus, Diogenes Street, Engomi 1516, Nicosia, Cyprus \label{euc}
    \and Scuola Normale Superiore, Piazza dei Cavalieri 7, 56126 Pisa, Italy\label{sns}
    \and Astrophysics Research Institute, Liverpool John Moores University, 146 Brownlow Hill, Liverpool L3 5RF, UK \label{ljmu}
    \and Kavli Institute for Cosmology, University of Cambridge, Madingley Road, Cambridge, CB3 0HA, UK \label{kic}
    \and Cavendish Laboratory, University of Cambridge, 19 JJ Thomson Avenue, Cambridge, CB3 0HE, UK \label{cambridge}
    \and Department of Astronomy, University of Maryland, College Park, Maryland, 20742 USA \label{uom}
    \and Joint Space-Science Institute, University of Maryland, College Park, Maryland, 20742 USA \label{jssi}
    }
    \date{Received ; accepted; }

 
  \abstract
  {We investigate the spatial extent and structure of the \cii line emission in a sample of 34 galaxies at $z=4-6$ from the \cii Resolved ISM in STar-forming galaxies with ALMA (CRISTAL) Survey. By modeling the distribution of the \cii line emission in the interferometric visibility data directly, we derive the effective radius of \cii line emission assuming an exponential profile. These measurements comprise not only isolated galaxies but also interacting systems that were identified thanks to the high spatial resolution of the data. The \cii line radius ranges from 0.5 to 3.5 kpc with an average value of $\langle R_{e,{\mathrm{[CII]}}}\rangle=1.90$ kpc. We compare the \cii sizes with the sizes of rest-frame ultraviolet (UV) and far-infrared (FIR) continua, which were measured from the HST F160W images and ALMA Band-7 continuum images, respectively. We confirm that the \cii line emission is more spatially extended than the continuum emission, with average size ratios of $\langle R_{e,{\rm [CII]}}/R_{e,{\rm UV}}\rangle=2.90$ and $\langle R_{e,{\rm [CII]}}/R_{e,{\rm FIR}}\rangle=1.54$, although about half of the FIR-detected sample shows a comparable spatial extent between the \cii line and the FIR continuum emission ($R_{e,{\rm [CII]}}\approx R_{e,{\rm FIR}}$).
  The residual visibility data of the best-fit model do not show statistical evidence of flux excess, indicating that the \cii line emission in star-forming galaxies can be characterized by an extended exponential disk profile. Overall, our results suggest that the spatial extent of \cii line emission can primarily be explained by photodissociation regions associated with star formation activity, while the contribution from diffuse neutral medium (atomic gas) and the effects of past merger events may further expand the \cii line distributions, causing their variations among our sample. {Finally, we report the negative correlation between the \cii surface density ($\Sigma_{\mathrm{[CII]}}$) and the Ly$\alpha$ equivalent width (EW$_{\mathrm{Ly}\alpha}$), and a possible negative correlation between $R_{e,{\rm [CII]}}/R_{e,{\rm UV}}$ and $\mathrm{EW}_{\mathrm{Ly}\alpha}$, which may be in line with the scenario that atomic gas component largely contributes to the extended \cii line emission.} Future three-dimensional analysis of Ly$\alpha$ and H$\alpha$ lines will shed light on the association of the extended \cii line emission with atomic gas and outflows.
  }

   \titlerunning{Spatial extent of \cii line emission in SFGs at $z=4-6$}
   \authorrunning{R. Ikeda et al.}
   \maketitle
%

\section{Introduction}

Understanding the interplay between gas, dust, and stars in galaxies in the early Universe is one of the biggest topics in modern extragalactic astronomy. In particular, the cold interstellar medium (ISM), which can be interpreted as a direct fuel for forming stars, is essential as it becomes the dominant contributor to the baryonic mass toward high redshift \citep{2020ARA&A..58..157T}. In this context, far-infrared (FIR) fine-structure lines have been known as an effective probe for star formation and the ISM of galaxies.

The \cii $^{2}P_{3/2}$$-$$^{2}P_{1/2}$ fine-structure line emission ($\nu_{\rm rest}=1900.537$ GHz, hereafter \cii line), emitted from ionized carbon (C$^{+}$), is one of the main coolants of the ISM among the FIR emission lines and has been the most popular tracer with which to study the cool gas component in high-redshift galaxies ($z\gtrsim4$; \citealp{2013ARA&A..51..105C}), as its redshifted millimeter wavelength is accessible through the atmospheric windows. Ionized carbon is excited by photoelectric heating of gas, whereby electrons are emitted from small dust grains or polycyclic aromatic hydrocarbons (PAHs) that absorb far-ultraviolet (FUV) photons from star-forming regions. Therefore, it has been suggested that \cii line emission can be used as a good tracer of photodissociation regions (PDRs; \citealp{1999RvMP...71..173H}; \citealp{2022ARA&A..60..247W}), as well as a star formation rate (SFR) indicator from nearby star-forming galaxies (SFGs; e.g., \citealp{1991ApJ...373..423S}; \citealp{2014A&A...570A.121P}; \citealp{2015ApJ...798...24K}; \citealp{2014A&A...568A..62D}; \citealp{2015ApJ...800....1H}; \citealp{2017ApJ...840...51L};  \citealp{2020ApJ...903...30B}; but see \citealp{2013ApJ...774...68D} for starburst galaxies) to high-redshift galaxies (e.g., \citealp{2015ApJ...813...36V}; \citealp{2018MNRAS.478.1170C}; \citealp{2020A&A...643A...3S}; \citealp{2024MNRAS.528..499L}). At the same time, \cii line emission is also used as a tracer of molecular gas (e.g., \citealp{1997ApJ...483..200M}; \citealp{2014A&A...561A.122L}; \citealp{2018MNRAS.481.1976Z}; \citealp{2020A&A...643A...5D}; \citealp{2020A&A...643A.141M}; \citealp{2022ApJ...929...92V}; \citealp{2024A&A...682A..24A}) and atomic gas (e.g., \citealp{2021ApJ...922..147H}; \citealp{2022ApJ...934L..27H}; \citealp{2022ApJ...939L...1V}), which may be compatible with the SFR tracer if the Kennicutt-Schmidt relation \citep{1998ApJ...498..541K} holds. {However, it has been known that the \cii line emission can originate from multiple phases of ISM beyond PDRs, such as ionized gas, making the interpretation of \cii line emission less straightforward.} With these potential capabilities, a large number of recent studies focusing on ISM conditions in high-redshift galaxies ($z\gtrsim4$) rely on observations of \cii line emission, thanks to the capability of the Atacama Large Millimeter/submillimeter Array (ALMA) and the Northern Extended Millimeter Array (NOEMA) interferometer.

If the majority of \cii line emission indeed originates from PDRs, one would expect its spatial extent in galaxies to also trace the region where active star formation takes place, which is bright in rest-frame ultraviolet (UV) continuum emission from young stellar populations. In contrast to this expectation, one remarkable feature of the \cii line emission in high-redshift galaxies is that they are more spatially extended than what is observable through optical/near-infrared wavelengths. 

Several studies have reported comparisons of spatial extent between \cii line emission and the rest-frame UV continuum observed by the Hubble Space Telescope (HST) for galaxies at $z>4$, both for individual galaxies (e.g., \citealp{2018MNRAS.478.1170C}; \citealp{2020ApJ...900....1F}) and for stacked images (e.g., \citealp{2022ApJ...934..144F}), providing the general picture that at high redshift the \cii sizes are larger than the UV sizes by a factor of more than two. Furthermore, \cite{2019ApJ...887..107F} report the detection of a 10 kpc scale ``\cii halo'' {with a beam-convolved radius of 10\,kpc ($R_{e,\mathrm{[CII]}}=5.6\pm0.1$\,kpc)}, which can be characterized by a two-component structure in the radial profile, by stacking 18 SFGs at $z=5-7$. With the detection of 75 \cii line emission from the ALPINE Survey (\citealp{2020A&A...643A...1L}; \citealp{2020A&A...643A...2B}; \citealp{2020ApJS..247...61F}), \cite{2020A&A...633A..90G} found evidence of the ``\cii halo'' through the stacking analysis and the broad spectral feature for a subsample of galaxies with higher SFRs, indicative of ongoing star-formation-driven outflows. {Although a visual classification based on morphology and kinematics was conducted \citep{2020A&A...643A...1L}, the studies referred to above cannot rule out the possibility that \cii line emission from close pairs or mergers is blended due to low spatial resolution ($\sim1''$).}

A handful of studies have also compared \cii line emission with the underlying FIR continuum emitted from dust grains heated by star formation or active galactic nuclei (AGNs; e.g., \citealp{2018ApJ...859...12G}; \citealp{2020ApJ...904..130V}; \citealp{2021A&A...654A..37D}; \citealp{2021ApJ...914...36I,2021ApJ...908..235I}; \citealp{2021ApJ...918...69U}; \citealp{2024A&A...686A.187P};  \citealp{2024ApJ...970....9B}; \citealp{2024ApJ...968....9W}), although these sources tend to be biased toward submillimeter galaxies or quasars due to their brightness in the FIR continuum. These studies show that the \cii line emission is comparable to or more extended than the FIR continuum emission, but the differences are not as significant as those for the rest-frame UV emission. Through the stacking analysis of 27 quasars at $z\gtrsim6$, \cite{2020ApJ...904..131N} found a comparable spatial extent between \cii line and FIR continuum emission extending up to $\sim10$\,kpc with no sign of broad line wings in stacked \cii line spectra.

Extended \cii line emission beyond the rest-frame UV and/or FIR continua directly implies the presence of metal-enriched gas up to a circumgalactic medium (CGM) scale even in the early Universe, and has been a matter of debate in recent years, given that cosmological zoom-in simulations could not reproduce such extended \cii line emission (\citealp{2019ApJ...887..107F}; \citealp{2020MNRAS.498.5541A}). Multiple scenarios have been proposed, such as satellite galaxies, extended star formation activity, cold streams, and outflows. Recent semi-analytical models support that the two-component halo structure in \cii line emission can be explained by cold outflows \citep{2020MNRAS.495..160P,2023MNRAS.519.4608P}, but none of the above scenarios are observationally conclusive yet. There are two reasons why it is difficult to elucidate the origin of such extended \cii line emission. First, \cii line emission can originate from multiple gas phases (ionized or neutral, atomic or molecular, PDR or H\,{\sc ii} regions, and shocked gas); thus, integrated \cii line emission is likely to trace a mixture of them. Second, previous \cii line observations were mostly performed with a spatial resolution of $\sim1''$ ($\sim6$\,kpc in a physical scale at $z=5$), hampering the measurements of accurate sizes and the study of morphology below this scale.

Therefore, to narrow down the origin of \cii line emission at high redshift, it is necessary to study galaxies at high resolution and compare them with other tracers. In fact, several studies have attempted to follow up the \cii line at high resolution with improved sensitivity, and have demonstrated that ALMA is capable of characterizing the \cii line in typical SFGs at $z=4-7$ down to a kiloparsec scale (\citealp{2021A&A...649A..31H}; \citealp{2022ApJ...934...64A}; \citealp{2023MNRAS.518.3183L}; \citealp{2023A&A...669A..46P}). Spatially resolved studies of \cii line emission have recently been reported for two systems (\citealp{2024A&A...689A.145S}; \citealp{2024arXiv240303379P}) as a part of the \cii Resolved ISM in STar-forming galaxies with ALma (CRISTAL) Survey (Herrera-Camus et al. in prep) \footnote{\url{https://sites.google.com/view/alma-cristal}}. \cite{2024A&A...689A.145S} discovered an elongated \cii line feature associated with the J1000+0234 system, composed of a dusty star-forming galaxy and a Lyman-break galaxy (CRISTAL-01a), while several scenarios (e.g., a conical outflow, a cold accretion stream) are possible for its origin. \cite{2024arXiv240303379P} analyzed CRISTAL-05, which was previously claimed as a ``\cii halo'' \citep{2020ApJ...900....1F}, and found a complex kinematical structure, implying that CRISTAL-05 is likely to be a merging system. Both studies exemplify the diverse characteristics of \cii line emission after the Epoch of Reionization and the necessity of high-resolution ALMA observations to understand their intrinsic morphology and structure. Nonetheless, each study stated above remains a case study, and it is difficult to draw general conclusions even by compiling these samples. A systemic census of deep and high-resolution \cii line observations with a larger and homogeneous sample is needed to provide a more general picture.

In this paper, we report on the size measurement and search for an extended halo component of the \cii line emission in SFGs at $z=4-6$. To analyze pure signals obtained from the interferometer without any processes of Fourier transform -- that is, synthesis imaging -- we focus on the analysis of  interferometric visibility data throughout this paper. We first describe the CRISTAL Survey, the galaxy sample, and the archival data that we used in this study in Section\,\ref{sec:Section2}. Section\,\ref{sec:Section3} presents the analysis of visibility modeling, including the classification based on multiplicity in either \cii line or rest-frame UV emission. We report the main results and discussions on the extended \cii line emission and correlation with other galaxy properties in Section\,\ref{sec:Section4} and summarize this study in Section\,\ref{sec:Section5}.

Throughout this paper, we assume a standard $\Lambda$CDM cosmology with $H_{0} = 70$\,km\,s$^{-1}$\,Mpc$^{-1}$, $\Omega_{\rm M}=0.3$, and $\Omega_{\Lambda}=0.7$. {We refer to the term ``\cii halo'' as a secondary structure which is more extended than the primary \cii line emission in the central regions.} We use the notation of $R_{e,{\rm X}}$ as a \textit {circularized} effective radius of a tracer X, a product of the major-axis effective radius, and a square root of the minor-to-major axis ratio (denotated as $q$), unless otherwise noted.

\section{Data} \label{sec:Section2}
\subsection{The CRISTAL Survey}
In this study, we use the data taken from the CRISTAL Survey, an ALMA Cycle 8 Large Program (Project ID: \#2021.1.00280.L, PI: R. Herrera-Camus). The observations were carried out during ALMA Cycle 8 and 9, and were completed in April 2023. The primary aim of the survey is to spatially resolve both \cii line and FIR continuum emission on $\sim1$\,kpc scales with ALMA Band\,7 observations for 19 star-forming main sequence galaxies at $z=4-6$. The targets were drawn from the ALPINE Survey \citep{2020A&A...643A...1L}. Targets were mass-selected with the selection of $\log(M_{\star}/M_{\odot})\geq9.5$ and were also required to have observations in the {HST/Wide Field Camera 3 (WFC3) F160W filter. These HST images were primarily obtained through observations of the CANDELS Survey \citep{2011ApJS..197...35G} as well as the COSMOS-DASH Survey (\citealp{2017PASP..129a5004M}; see more details in \citealp{2020ApJS..247...61F} and Herrera-Camus et al. in prep).} Here, we summarize aspects of the ALMA observations that are relevant to the analysis of this study. 

The combination of multi-configuration observations is essential to achieve both sufficiently large maximum recoverable scale (MRS) and high spatial resolution. Therefore, each target was observed {using multiple configurations} of the ALMA 12m array, one with compact (C43-1/2/3) and another with extended (C43-4/5/6) array configuration. The former covers low spatial frequencies ($uv \, {\rm distance}\lesssim150$\,-\,$300$\,k$\lambda$)\,\footnote{We use a term of `$uv$ distance' as a radial distance of visibility data in the ($u,v$) plane in units of k$\lambda$, defined as $uv \, {\rm distance}=\sqrt{u^{2}+v^{2}}$.} in the visibility plane that are sensitive to large-scale structure {($\theta_{\mathrm{beam}}=0\farcs29$\,-\,$1\farcs00, \theta_{\mathrm{MRS}}=5\farcs28$\,-\,$9\farcs79$)}, while the latter covers high spatial frequencies ($uv \, {\rm distance}\gtrsim350$\,-\,$1500$\,k$\lambda$, depending on the target) that are sensitive to small-scale structure ($\theta_{\mathrm{beam}}=0\farcs07$\,-\,$0\farcs35, \theta_{\mathrm{MRS}}=1\farcs15$\,-\,$5\farcs15$).

We used the ALMA correlators of the 1.875 GHz width spectral windows (SPWs) with the frequency division mode (FDM). The observing \cii frequency for each object was adjusted in the middle of two adjacent SPWs in the upper sideband unless the \cii line emission was not detected in the ALPINE Survey \citep{2020A&A...643A...2B}.

\subsection{ALMA archive}
\subsubsection{The ALPINE Survey}
To search for faint extended emission, deep integration in the {more} compact array configuration, covering short spatial frequencies, is particularly important. Therefore, we combined the data from the ALPINE Survey {(\#2017.1.00428.L)} with the CRISTAL data to achieve better sensitivity to the extended structure.

Contrary to the CRISTAL Survey, the ALPINE Survey adopts the Time Division Mode for the SPW setup, in which the edges of 2~GHz bandwidth are severely affected by low power. We find that in three sources the \cii lines spread across the two adjacent SPWs with insufficient overlapping frequencies (CRISTAL-06, CRISTAL-08, and CRISTAL-23c), and therefore for these fields we decided not to combine the ALPINE data in our analysis. Also, we did not include the ALPINE data of CRISTAL-07c, a serendipitously detected galaxy at $z=5.16$, as it is located at the edge of the Band-7 field of view (FoV) with the primary beam gain level of $\sim0.3$, while in the CRISTAL Survey, the FoV is adjusted to have comparable sensitivity in both CRISTAL-07a/07b and CRISTAL-07c.

\begin{figure}[ht!]
    \centering
	\includegraphics[width=0.85\columnwidth]{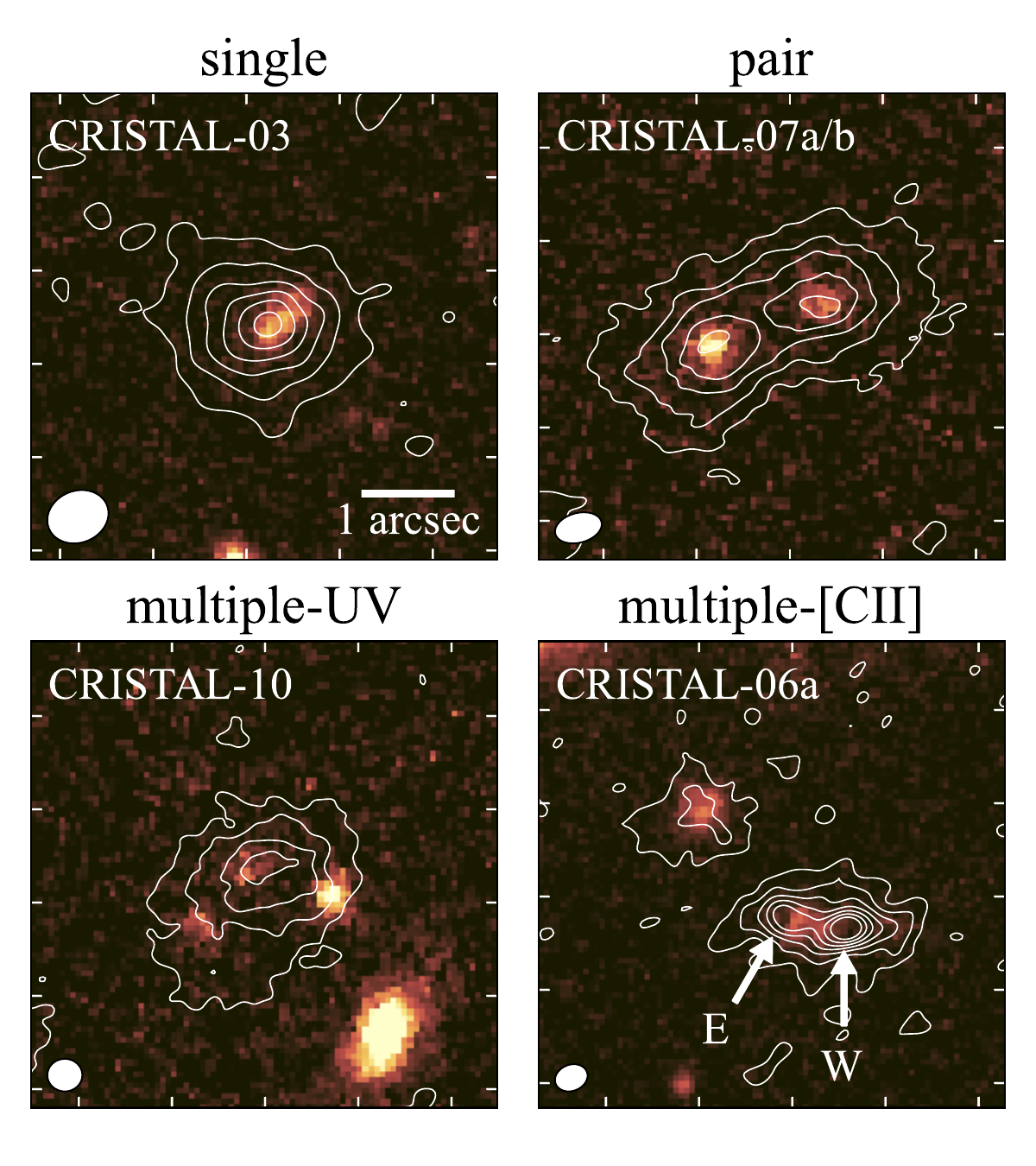}
    \caption[aaa]{Classification of the galaxies based on a multiplicity of \cii line (with contours) and rest-frame UV (background, HST/WFC3 F160W) emission. A $5''\times5''$ region is shown in each panel. The contours start at $2\sigma$ and increase in steps of $3\sigma$ until $20\sigma$. The Gaussian FWHM fit from the synthesized dirty beam is shown as a white ellipse in the lower-left corner of each panel.{\footnotemark} Two arrows shown in a panel of multiple-[C\,{\sc ii}] (CRISTAL-06a) correspond to two \cii line components (denoted as `W' and `E') that are visually identified.}
    \label{fig:Fig1}
\end{figure}

\subsubsection{The CRISTAL+ Sample}
In addition to the main targets of the CRISTAL survey, there exist several programs that used ALMA to observe \cii line emission with similar resolutions and sensitivities. This includes six galaxies: HZ4, HZ7, HZ10, DEIMOS\_COSMOS\_818760, DEIMOS\_COSMOS\_873756, and vuds\_cosmos\_8326. We have included these archival data as the supplementary sample of the CRISTAL Survey and hereafter refer to these six targets as CRISTAL-20 to CRISTAL-25 (Herrera-Camus et al. in prep). The \cii line emission of the first three was initially reported by \cite{2015Natur.522..455C} and followed up by individual projects for high-resolution data {(\#2018.1.01605.S; \citealp{2021A&A...649A..31H},  \#2018.1.01359.S; \citealp{2023MNRAS.518.3183L}, \#2019.1.01075.S; \citealp{2024A&A...691A.133V} and \citealp{2024arXiv241109033T})}. CRISTAL-20 was also observed in the ALPINE Survey as DEIMOS\_COSMOS\_494057. The latter three are the three brightest objects in \cii line emission among the sample of the ALPINE Survey \citep{2020A&A...643A...2B} and followed up at higher resolution {under the ALMA Cycle 7 program (\#2019.1.00226.S; \citealp{2024A&A...686A.156D})}. We combine the data taken by the ALPINE Survey with these follow-up observations, similar to the CRISTAL main targets.

In summary, the data set of this study comprises 24 ALMA Band-7 fields, containing 35 main-sequence SFGs and two dusty star-forming galaxies (DSFGs) detected in \cii line at $z=4-6$. We excluded CRISTAL-18 from the analysis because we do not detect \cii line emission from this source. {We present the list of our sample cross-matched to the other names presented in previous studies in Table\,\ref{table:TableA1}.}

\footnotetext{{However, we note that a synthesized dirty beam created using multiple array configurations frequently deviates from a single Gaussian profile and exhibits an elongated feature toward larger scales. Thus, we caution the reader that most of the dirty images presented in this paper are convolved with a point spread function (PSF) elongated than a single Gaussian.}}

\section{Analysis} \label{sec:Section3}

\subsection{Classification based on multiplicity} \label{subsec:Section3.1}

By virtue of a kiloparsec-scale resolution in \cii line emission and ancillary multi-wavelength data of our sample, we can compare the spatial extent between \cii line and UV continuum emission including complex systems such as interacting galaxies. One possible origin of extended \cii line emission is the presence of tidal features or gas removal caused by mergers (\citealp{2024A&A...689A.145S}; \citealp{2024arXiv240303379P}). Therefore, we first classified the sample based on whether each galaxy is likely to be physically associated with other galaxies. \cite{2024A&A...690A.197M} classified the CRISTAL sample into single and multiple sources solely based on the multiplicity in HST/WFC3 F160W images. In this study, we take a step forward to add information from the \cii line distribution to visually classify the CRISTAL sample into four categories:
\begin{itemize}
  \item[(a)] single - an isolated galaxy with a single component of the \cii line emission. We classify 17 sources ($\sim50$\,\%) of the CRISTAL sample as single. 
  \item[(b)] pair - a pair of \cii line detections with a projected angular separation of less than $\sim3$ arcsec. The velocity offset among the pair is confirmed to be small (<300 km/s) so that there is an overlap of \cii line frequencies. We classify nine sources ($\sim30$\,\%) as pair.
  \item[(c)] multiple-UV - a system comprising of multiple compact blobs in the UV continuum with overlapping extended \cii line emission. Such systems are likely to be complex merging systems {instead of unevenly dust-obscured systems, since their FIR continuum emission is either faint (S/N$<5$) or entirely overlapping the UV blobs \citep{2024A&A...690A.197M}}. The difference from the pair systems is that the \cii line emission cannot be resolved into each UV counterpart with the existing data. We classified five sources (CRISTAL-01a, CRISTAL-01b, CRISTAL-02, CRISTAL-10, and CRISTAL-13) into this category. 
  \item[(d)] multiple-[C\,{\sc ii}] - a system in which multiple peaks in \cii line emission can be found around a single UV component. Only CRISTAL-06a exhibits this feature.
\end{itemize}

Figure~\ref{fig:Fig1} summarizes our classification, showcasing examples of the \cii line and rest-frame UV distributions for each class. We note, however, that this is still a crude classification, as we do not consider the dynamical state of each galaxy obtained from kinematical analysis. For example, we classify CRISTAL-05 as single from its appearance. A detailed kinematical analysis of the \cii line revealed that it is actually a complex galaxy pair projected along the line of sight \citep{2024arXiv240303379P}. 

We also find that three galaxies (CRISTAL-11, CRISTAL-15, and CRISTAL-17) classified as single have pair-like morphology when looked with sharper {images taken by James Webb Space Telescope (JWST; \citealp{2024arXiv240910963L})}. However, the resolution of the \cii line emission is coarser than the separation of two clumps identified by JWST, and thus it would be difficult to characterize their sizes and perform the one-to-one comparison. We thus decide to rely on the classification based on rest-frame UV morphology primarily due to the data homogeneity, and we note that the discussions and conclusions we present in the following would not significantly be affected by the choice of our classification, because the sizes of the continuum emission are not constrained for two out of these three galaxies (Table\,\ref{table:Table2}).

\begin{figure}
    \centering
	\includegraphics[width=0.98\columnwidth]{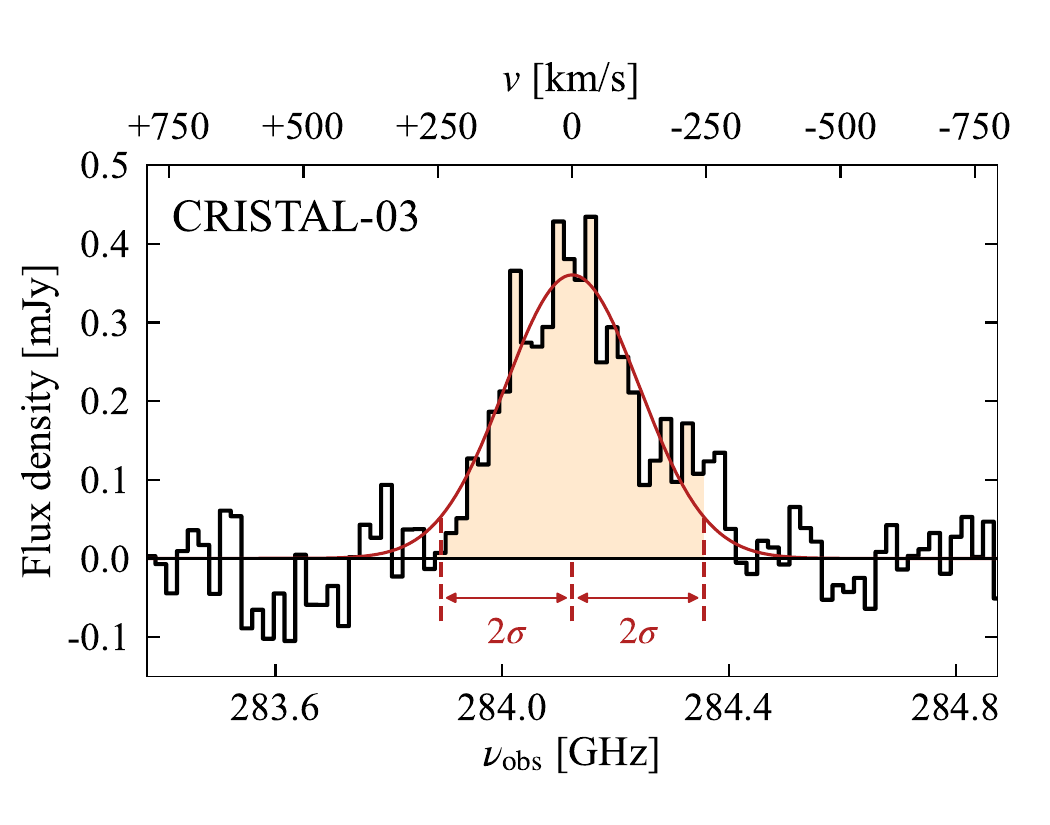}
    \caption{\cii line spectrum of CRISTAL-03 extracted from a 1-arcsec aperture. The best-fit Gaussian is shown as a solid red curve. We bin the frequency range of the filled region for the \cii size measurements, which corresponds to $\pm2\sigma$ from the central frequency.}
    \label{fig:Fig2}
\end{figure}

\begin{figure*}[h!]
    \centering
    \includegraphics[width=\linewidth]{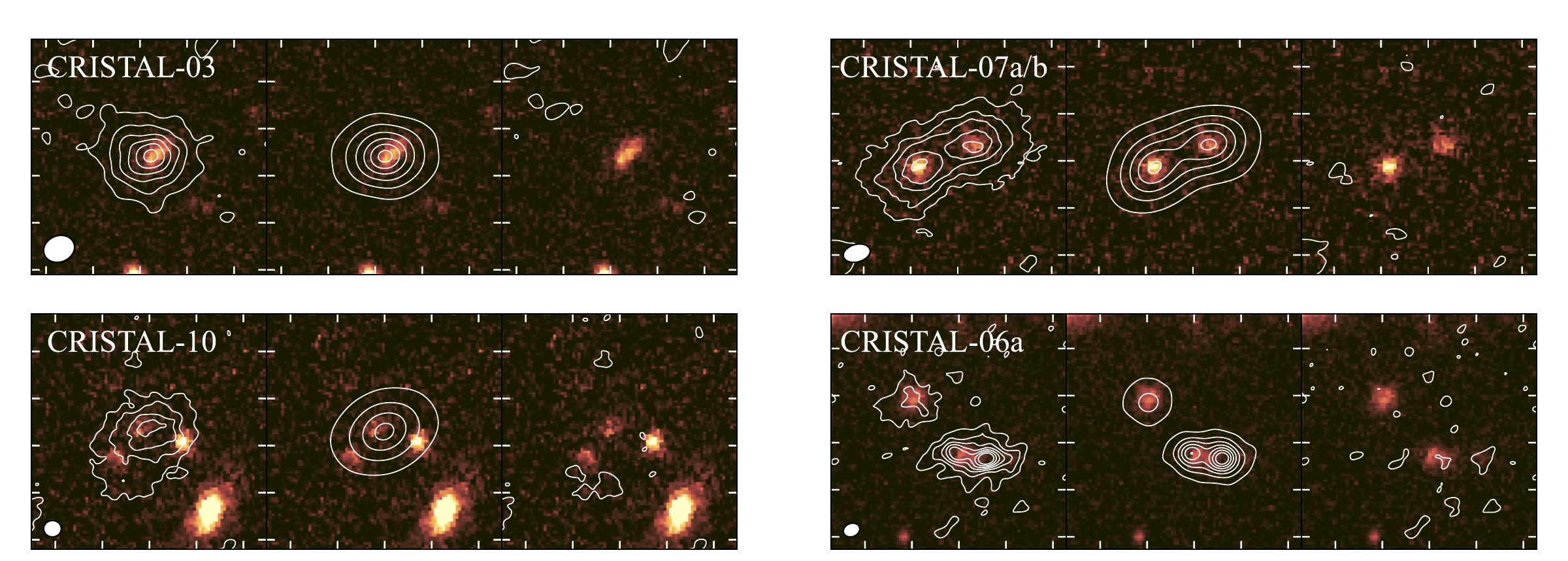}
    \caption{{Visualization of the visibility modeling of the \cii line emission (white contours) in four systems that we showed in Figure\,\ref{fig:Fig1}. The left panel shows the dirty image of a single cube used for the visibility modeling (Section\,\ref{subsec:Section3.2}). {The Gaussian FWHM fit from the dirty beam is shown as a white ellipse in the lower-left corner.} The middle and right panel show the best-fit exponential disk model and the residual created by subtracting the model from the data, respectively. The contours start at $2\sigma$ and increase in steps of $3\sigma$ until $20\sigma$. The residual maps do not show any peaks above the $2\sigma$ level around the rest-frame UV counterpart, which indicates that the observed emission is characterized by the best-fit model well. The rest of the sample is shown in Figure\,\ref{fig:FigA1} and \ref{fig:FigA2}.}}
    \label{fig:Fig3}
\end{figure*}

\subsection{{Size measurements} \label{subsec:Section3.2}} 

\subsubsection{{\cii line emission}} \label{subsec:Section3.2.1}
The data from the CRISTAL Survey with a kiloparsec-scale resolution allows us to robustly quantify the sizes of \cii line emission. While the majority of the sample has a sufficient signal-to-noise ratio ($\mathrm{S/N}$) to perform size measurements in the image domain, several sources have a moderate S/N of $5\lesssim\mathrm{S/N}\lesssim10$. In such cases, the measured sizes in the image domain could be overestimated, because either the best-fit model takes noise into account, or the smearing effect of the clean beam affects the fittings (\citealp{2020ApJ...901...74T}; \citealp{2024A&A...690A.197M}). {In addition, the images created through the CLEAN algorithm may overestimate the flux densities outside the CLEAN masks due to the deviation of the synthesized dirty beam from a single Gaussian. To avoid these issues}, we consistently perform our \cii size measurements by modeling interferometric visibility data. The visibility analysis is also favorable assessing faint extended features in \cii line emission, as the imaging procedure requires several parameters to be considered (e.g., weighting of visibility data, threshold, size of the masked region for the CLEAN algorithm) and therefore complicates the optimization of the best parameter sets \citep{2023MNRAS.518..691J}.  

As a first step of the \cii size measurements, we define {the velocity range ($\Delta v$)} of \cii line emission. To perform a quantitatively fair comparison of \cii sizes among our sample in terms of velocity width range, we took the following steps to define the line width of the \cii line emission and create a form of visibility data optimized for size measurements: (1) extract spectrum with a 1-arcsec aperture from the data cube (20 km/s width per channel). Here, we used the data cube of the \cii line emission, produced by the {\tt CASA/tclean} task \citep{2022PASP..134k4501C}. The {\tt tclean} task was performed with {\tt auto-multithresh} algorithm and cleaned down to $1\sigma$ (Herrera-Camus et al. in prep). We adopted the robust parameter of either Briggs ($R=0.5$) or Natural ($R=2.0$) weighting that creates the synthesized beam size of $\theta_{\rm beam}=0\farcs3$\,--\,$0\farcs5$, (2) fit the extracted spectrum with a single Gaussian $f(\nu)\propto \exp[(\nu-\nu_{0})^{2}/2\sigma^{2}]$ using the {\tt curve\_fit} function within the {\tt scipy} package \citep{2020NatMe..17..261V}. From the best-fit Gaussian function, we adopted the spectral range that contains 95\,\% ($=$\,[2.5\,\%, 97.5\,\%]) of the total \cii line flux, corresponding to $4\sigma$ ($\pm2\sigma$ from the central frequency $\nu_{0}$) width of Gaussian ($\approx1.7$ Gaussian FWHM). We show a spectrum of CRISTAL-03 as an example in Figure\,\ref{fig:Fig2}. The $\pm2\sigma$ velocity range includes almost all of the \cii line emission, which assures that the \cii size measurements will be performed by fitting the representative velocity range of the emission. (3) Finally, we created a visibility measurement set that contains only information on the spectral range calculated above. {Here, we used the visibility data where the FIR continuum emission was subtracted by using {\tt CASA/uvcontsub} task.} The fitting of continuum slope was performed in the spectral range avoiding the \cii frequency of each object. We then binned the visibilities into a single channel by specifying {\tt nchan = 1} in the {\tt CASA/mstransform} task. As an exceptional case, we fit the \cii spectrum of CRISTAL-22a, which shows a clear double-peaked line profile,  with a double Gaussian. This exceptionally broad and complex \cii spectrum of CRISTAL-22a is likely due to a combination of merger and complex kinematical structure of the central component of CRISTAL-22a, which are verified from both [O\,{\sc iii}]\,5007\AA\,line emission map \citep{2024arXiv240512955J} and detailed kinematical analysis of the \cii line emission \citep{2024arXiv241109033T}.

\begin{figure}
    \centering
	\includegraphics[width=0.95\columnwidth]{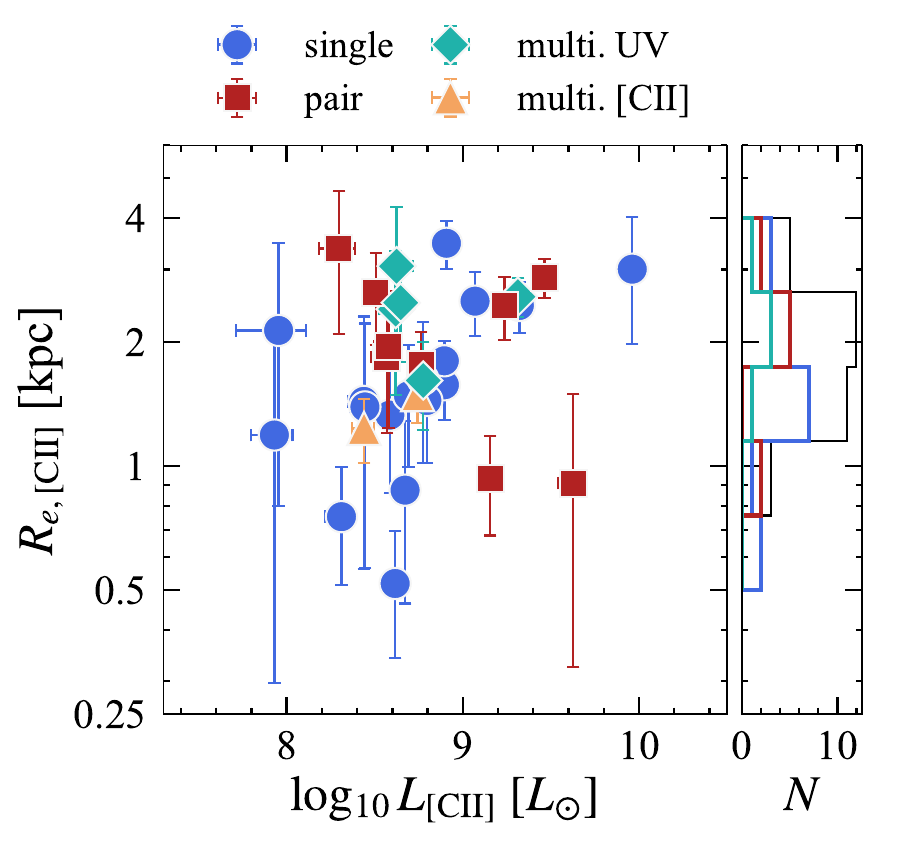}
    \caption{\cii luminosity-size relation of the CRISTAL sample. Symbols are based on the classification defined in Section \ref{subsec:Section3.1}. {Each component of the pair/multiple-[C\,{\sc ii}] systems are shown separately.} The right panel shows the histogram of \cii sizes among categories with matched colors. We omit a histogram of multiple-[C\,{\sc ii}] because of the small sample size ($N=2$). A histogram with a solid black line shows the size distribution of the entire CRISTAL sample. {The \cii sizes show the diversity at fixed \cii luminosity, in which relatively compact sizes ($R_{e,\mathrm{[CII]}}\lesssim1$\,kpc) were challenging to be measured in the previous low-resolution data.}}
    \label{fig:Fig4}
\end{figure}


We used the {\tt UVMULTIFIT} \citep{2014A&A...563A.136M} to model the visibilities of the \cii line emission in the CRISTAL sample. The {\tt UVMULTIFIT} implements simultaneous fitting of multiple components, which is useful for characterizing pair galaxies. We assumed a two-dimensional (2-D) exponential disk model ($n=1$) for all of the \cii components and first allow six parameters (R.A., Decl., total flux density, major-axis FWHM, minor-to-major axis ratio, and position angle of major axis) free. 
Some of the fittings fail to converge when the peak S/N of the \cii line emission is not high enough (typically $\mathrm{S/N}<8$). In this case, we fixed the projected minor-to-major axis ratio to unity. 

The fit components were determined based on the classification described in Section \ref{subsec:Section3.1}. For single and multiple-UV classes, we adopt one-component modeling, and otherwise, we adopt two-component modeling. {We show the images of the visibility data, the best-fit model, and the residual in Figure\,\ref{fig:Fig3} for four systems that we showed in Figure\,\ref{fig:Fig1}. We show the rest of the sample in Figure\,\ref{fig:FigA1} and \ref{fig:FigA2}. As can be seen in Figure\,\ref{fig:Fig3}, the best-fit model characterizes the observed \cii line emission well and no signal greater than $2\sigma$ significance level is present in the central $2''$-region, securing the validity of the fitting.} Our analysis yielded structural parameters of a total number of 34 \cii line components, including J1000+0234 \citep{2008ApJ...681L..53C}, a massive submillimeter galaxy ($S_{870\mu{\rm m}}=8.03$ mJy; \citealp{2024A&A...689A.145S}) located in the proximity to CRISTAL-01a.

Finally, we convert the fitting outputs of the total flux density and major-axis FWHM into the total \cii luminosity ($L_{\mathrm{[CII]}}$) and the effective radius ($R_{e,\mathrm{[CII]}}$), respectively. For a 2-D exponential disk model, the major axis FWHM and major axis effective radius can be related as $R_{e,\mathrm{major}}=1.21\times\mathrm{FWHM}$, therefore we obtained a circularized effective radius by calculating $R_{e,\mathrm{[CII]}}=1.21\times\mathrm{FWHM}\sqrt{q}$, where $q$ is the minor-to-major axis ratio. We estimate the uncertainty by using error propagation. For the uncertainty of each parameter (FWHM and $q$), we adopt the output from the {\tt UVMULTIFIT}, which is estimated from the post-fit covariance matrix \citep{2014A&A...563A.136M}. We summarize our fitting results in Table\,\ref{table:Table1}. We calculate an average value and standard deviation of the \cii sizes of $\langle R_{e,\mathrm{[CII]}}\rangle=1.90\pm0.77$\,kpc. To fairly compare with the fitting results from the low-resolution archival data alone, we iterate the above procedure by only using the data taken in the ALPINE Survey. We present the results and discussion in Appendix \ref{sec:AppendixC}. In brief, we find a good agreement  between two measurements for the \cii fluxes and sizes ($R_{e,\mathrm{[CII]}}\gtrsim1$\,kpc). On the other hand, we find that the axis ratios tend to be poorly constrained when only the ALPINE data is used.
 
In Figure\,\ref{fig:Fig4}, we show a \cii luminosity-size relation of the CRISTAL sample. To see the dependence of multiplicity, symbols are distinguished based on the classification presented in Section\,\ref{subsec:Section3.1}. Although we expect to find larger \cii sizes in pair and multiple-UV systems compared to single systems, we were unable to statistically distinguish their size distributions by performing the Kolmogorov-Smirnov test (with $p$\,values of 0.23 and 0.12 for the pair and multiple-UV, respectively), largely due to our limited sample size. Overall, Figure\,\ref{fig:Fig4} showcases the diversity of \cii sizes at fixed \cii luminosity, and we find that low-resolution data used in previous studies were unable to measure relatively compact sizes ($R_{e,\mathrm{[CII]}}\lesssim1$\,kpc).

{\subsubsection{Rest-frame UV and FIR continuum emission} \label{subsec:Section3.2.2}}

To evaluate the spatial extent of \cii line emission, we refer to two continuum size measurements (observing wavelengths of 1.6\,$\mu$m and 870\,$\mu$m), which were both performed in \cite{2024A&A...690A.197M}. The rest-frame UV sizes ($R_{e,{\rm UV}}$) are measured with {\tt GALFIT} \citep{2002AJ....124..266P} using the HST/WFC3 F160W images. A single S\'{e}rsic model with an index of $n=1$ is assumed (except for CRISTAL-13), and nearby objects seen in the HST images are masked by using {\tt SExtractor} \citep{1996A&AS..117..393B}. The rest-frame FIR sizes ($R_{e,\mathrm{FIR}}$) are measured from {the ALMA Band-7 continuum visibilities} with the same method using {\tt UVMULTIFIT} described in Section\,\ref{subsec:Section3.2.1}.

{\cite{2024A&A...690A.197M} classify the CRISTAL galaxies into either single or multiple systems based on the multiplicity in the HST/WFC3 F160W images. The multiple systems correspond to either pair, multiple-UV, or multiple-[C\,{\sc ii}] classified in Section\,\ref{subsec:Section3.1} of this study. Unlike \cii line emission, all of the FIR continuum emission was fit in a single exponential disk component due to the lower S/N detection compared to the \cii line emission.  While this hampers the fair comparison between FIR and \cii line sizes for multiple-UV, or multiple-[C\,{\sc ii}] systems, we find that the FIR continuum is detected in either one of the pair systems (e.g., CRISTAL-04a, CRISTAL-07a), and hence we adopt the FIR sizes of these galaxies in this study.}

\begin{table*}
\renewcommand{\arraystretch}{1.4}
\caption{Structural parameters of \cii line emission in the CRISTAL galaxies.}
\label{table:Table1}
\centering
  \begin{tabular}{lrrcccccc}
    \hline\hline 
    ID & R.A. & Decl. & $I_{\rm [CII]}$ & {$\Delta v$} ${^{(a)}}$ & $R_{{e,\rm[CII]}}$ & $q_{\rm [CII]}$ & PA$_{\rm [CII]}$ & $\log_{10}(L_{\rm[CII]}/L_{\odot})$ \\
     & & & (Jy~km/s) & {(km/s)} & (kpc) & & (deg) &  \\
    \hline
CRISTAL-01a & 10:00:54.508 & 2:34:34.46 & $0.68\pm0.16$ & 789 & $3.05\pm1.19$ & $0.59\pm0.27$ & $21\pm24$ & $8.62^{+0.09}_{-0.12}$  \\
CRISTAL-01b & 10:00:54.778 & 2:34:28.27 & $0.67\pm0.13$ & 376 & $2.40\pm0.92$ & $0.44\pm0.22$ & $67\pm15$ & $8.62^{+0.08}_{-0.09}$  \\
CRISTAL-02 & 10:00:21.499 & 2:35:11.01 & $2.63\pm0.16$ & 533 & $2.58\pm0.28$ & $0.52\pm0.07$ & $153\pm5$ & $9.31^{+0.03}_{-0.03}$  \\
CRISTAL-03 & 9:59:53.254 & 2:07:05.40 & $0.44\pm0.05$ & 490 & $1.33\pm0.47$ & $0.81\pm0.34$ & $90\pm54$ & $8.59^{+0.04}_{-0.05}$  \\
CRISTAL-04a & 9:58:57.907 & 2:04:51.36 & $0.94\pm0.07$ & 308 & $1.76\pm0.36$ & $0.90\pm0.21$ & $50\pm66$ & $8.76^{+0.03}_{-0.04}$  \\
CRISTAL-04b & 9:58:57.928 & 2:04:52.85 & $0.32\pm0.07$ & 129 & $3.37\pm1.27$ & $0.74\pm0.31$ & $57\pm42$ & $8.30^{+0.09}_{-0.12}$  \\
CRISTAL-05 & 10:00:09.422 & 2:20:13.84 & $0.94\pm0.08$ & 523 & $1.57\pm0.28$ & $0.31\pm0.08$ & $160\pm5$ & $8.90^{+0.03}_{-0.04}$  \\
CRISTAL-06a W & 10:01:00.894 & 1:48:33.66 & $0.89\pm0.07$ & 298 & $1.49\pm0.22$ & $0.50\pm0.08$ & $94\pm6$ & $8.74^{+0.03}_{-0.03}$  \\
CRISTAL-06a E & 10:01:00.940 & 1:48:33.81 & $0.44\pm0.06$ & 298 & $1.23\pm0.22$ & -- & -- & $8.44^{+0.06}_{-0.07}$  \\
CRISTAL-06b & 10:01:00.999 & 1:48:34.85 & $0.52\pm0.07$ & 332 & $2.63\pm0.66$ & $0.71\pm0.20$ & $123\pm23$ & $8.51^{+0.06}_{-0.06}$  \\
CRISTAL-07a & 10:00:04.058 & 2:37:35.84 & $0.49\pm0.09$ & 325 & $1.84\pm0.64$ & $0.89\pm0.39$ & $0\pm13$ & $8.57^{+0.07}_{-0.09}$  \\
CRISTAL-07b & 10:00:03.974 & 2:37:36.36 & $0.50\pm0.08$ & 351 & $1.95\pm0.71$ & $0.73\pm0.31$ & $67\pm40$ & $8.58^{+0.07}_{-0.08}$  \\
CRISTAL-07c & 10:00:03.223 & 2:37:37.75 & $1.56\pm0.15$ & 413 & $2.51\pm0.45$ & $0.74\pm0.15$ & $70\pm19$ & $9.07^{+0.04}_{-0.04}$  \\
CRISTAL-08 & 3:32:19.044 & -27:52:38.21 & $1.35\pm0.12$ & 320 & $3.47\pm0.47$ & $0.74\pm0.12$ & $8\pm15$ & $8.91^{+0.04}_{-0.04}$  \\
CRISTAL-09 & 9:59:00.893 & 2:05:27.57 & $0.56\pm0.07$ & 518 & $0.87\pm0.41$ & $0.78\pm0.46$ & $13\pm66$ & $8.67^{+0.05}_{-0.06}$  \\
CRISTAL-10 & 10:02:04.134 & 1:55:44.30 & $0.51\pm0.07$ & 386 & $2.49\pm0.71$ & $0.73\pm0.23$ & $122\pm29$ & $8.65^{+0.06}_{-0.07}$  \\
CRISTAL-11 & 10:00:32.598 & 2:15:28.44 & $1.05\pm0.11$ & 529 & $1.44\pm0.43$ & $0.65\pm0.23$ & $105\pm23$ & $8.80^{+0.04}_{-0.05}$  \\
CRISTAL-12 & 3:32:11.943 & -27:41:57.44 & $0.10\pm0.03$ & 224 & $1.19\pm0.89$ & $0.53\pm0.54$ & $18\pm44$ & $7.93^{+0.10}_{-0.14}$  \\
CRISTAL-13 & 10:00:41.171 & 2:17:14.24 & $0.95\pm0.09$ & 382 & $1.61\pm0.39$ & $0.47\pm0.15$ & $63\pm11$ & $8.78^{+0.04}_{-0.04}$  \\
CRISTAL-14 & 9:59:47.068 & 2:22:32.92 & $0.35\pm0.07$ & 534 & $0.75\pm0.24$ & -- & -- & $8.31^{+0.08}_{-0.10}$  \\
CRISTAL-15 & 10:00:47.657 & 2:18:02.09 & $0.44\pm0.08$ & 540 & $1.43\pm0.87$ & $0.47\pm0.27$ & $116\pm20$ & $8.44^{+0.07}_{-0.09}$  \\
CRISTAL-16 & 3:32:15.899 & -27:41:24.36 & $0.33\pm0.06$ & 445 & $1.39\pm0.83$ & $0.17\pm0.13$ & $68\pm7$ & $8.44^{+0.07}_{-0.08}$  \\
CRISTAL-17 & 10:00:39.132 & 2:25:32.66 & $0.10\pm0.04$ & 312 & $2.13\pm1.33$ & -- & -- & $7.95^{+0.15}_{-0.24}$  \\
CRISTAL-19 & 10:00:05.105 & 2:03:12.10 & $0.64\pm0.05$ & 497 & $1.48\pm0.48$ & $0.86\pm0.24$ & $158\pm50$ & $8.69^{+0.03}_{-0.04}$  \\
CRISTAL-20 & 9:58:28.503 & 2:03:06.56 & $0.94\pm0.04$ & 330 & $1.80\pm0.21$ & $0.57\pm0.07$ & $25\pm6$ & $8.89^{+0.02}_{-0.02}$  \\
CRISTAL-21 & 9:59:30.466 & 2:08:02.62 & $0.77\pm0.12$ & 422 & $1.62\pm0.61$ & $0.46\pm0.17$ & $76\pm12$ & $8.77^{+0.06}_{-0.07}$  \\
CRISTAL-22a & 10:00:59.291 & 1:33:19.43 & $4.91\pm0.87$ & 1032 & $0.91\pm0.58$ & $0.40\pm0.22$ & $107\pm14$ & $9.63^{+0.07}_{-0.08}$  \\
CRISTAL-22b & 10:00:59.250 & 1:33:19.42 & $1.66\pm0.25$ & 924 & $0.93\pm0.25$ & $0.49\pm0.15$ & $80\pm11$ & $9.16^{+0.06}_{-0.07}$  \\
CRISTAL-23a & 10:01:54.863 & 2:32:31.52 & $4.66\pm0.27$ & 462 & $2.87\pm0.31$ & $0.76\pm0.08$ & $113\pm11$ & $9.46^{+0.02}_{-0.03}$  \\
CRISTAL-23b & 10:01:54.970 & 2:32:31.51 & $2.77\pm0.27$ & 442 & $2.45\pm0.43$ & $0.82\pm0.15$ & $123\pm28$ & $9.24^{+0.04}_{-0.04}$  \\
CRISTAL-23c & 10:01:54.682 & 2:32:31.42 & $0.66\pm0.11$ & 465 & $0.52\pm0.18$ & -- & -- & $8.62^{+0.07}_{-0.08}$  \\
CRISTAL-24 & 10:00:02.715 & 2:37:39.99 & $14.76\pm2.50$ & 1291 & $3.00\pm1.03$ & $0.74\pm0.25$ & $92\pm31$ & $9.96^{+0.07}_{-0.08}$  \\
CRISTAL-25 & 10:01:12.502 & 2:18:52.55 & $3.35\pm0.19$ & 477 & $2.45\pm0.35$ & $0.65\pm0.10$ & $35\pm10$ & $9.32^{+0.02}_{-0.03}$  \\
\hline
  \end{tabular}
  \tablefoot{{\,($a$)\, Velocity range containing 95\,\% ($=$\,[2.5\,\%, 97.5\,\%]) of the total \cii line flux. To convert to the 80\,\% ($=$\,[10\,\%, 90\,\%]) velocity range, it is necessary to apply a multiplication factor of 0.654.}}
\end{table*}

\begin{table*}
\renewcommand{\arraystretch}{1.4}
\caption{Effective radii and size ratios in continuum emission.}
\label{table:Table2}
\centering
  \begin{tabular}{lccccc}
    \hline\hline 
    ID & $R_{e,{\rm UV}}$ & $R_{e,{\rm FIR}}$ & $R_{e,{\rm [CII]}}/R_{e,{\rm UV}}$ & $R_{e,{\rm [CII]}}/R_{e,{\rm FIR}}$ & Class. \\
    & (kpc) & (kpc) & & & \\
    \hline
     CRISTAL-01a & $1.02\pm0.06$ & -- & $3.0\pm1.2$ & -- & multi-UV \\
CRISTAL-01b & $0.63\pm0.05$ & $2.18\pm1.01$ & $3.8\pm1.5$ & $1.1\pm0.7$ & multi-UV \\
CRISTAL-02 & -- & $1.99\pm0.46$ & -- & $1.3\pm0.3$ & multi-UV \\
CRISTAL-03 & $1.02\pm0.05$ & -- & $1.3\pm0.5$ & -- & single \\
CRISTAL-04a & $1.01\pm0.05$ & $0.86\pm0.47$ & $1.7\pm0.4$ & $2.1\pm1.2$ & pair \\
CRISTAL-04b & $0.96\pm0.14$ & -- & $3.5\pm1.4$ & -- & pair \\
CRISTAL-05 & $0.36\pm0.04$ & $1.45\pm0.63$ & $4.3\pm0.9$ & $1.1\pm0.5$ & single \\
CRISTAL-06a\,W & -- & $2.03\pm0.47$ & -- & $0.7\pm0.2$ & multi-\cii \\
CRISTAL-06a\,E & -- & -- & -- & -- & multi-\cii \\
CRISTAL-06b & $1.25\pm0.07$ & -- & $2.1\pm0.5$ & -- & pair \\
CRISTAL-07a & $0.90\pm0.06$ & $1.24\pm0.95$ & $2.0\pm0.7$ & $1.5\pm1.2$ & pair \\
CRISTAL-07b & $1.51\pm0.13$ & -- & $1.3\pm0.5$ & -- & pair \\
CRISTAL-07c & $0.88\pm0.11$ & $0.86\pm0.48$ & $2.8\pm0.6$ & $2.9\pm1.7$ & single \\
CRISTAL-08 & $2.53\pm0.09$ & -- & $1.4\pm0.2$ & -- & single \\
CRISTAL-09 & $0.17\pm0.05$ & $0.62\pm0.47$ & $5.2\pm3.0$ & $1.4\pm1.3$ & single \\
CRISTAL-10 & -- & $2.38\pm1.33$ & -- & $1.0\pm0.7$ & multi-UV \\
CRISTAL-11 & $0.44\pm0.16$ & $0.77\pm0.60$ & $3.3\pm1.5$ & $1.9\pm1.6$ & single \\
CRISTAL-12 & -- & -- & -- & -- & single \\
CRISTAL-13 & -- & $0.60\pm0.46$ & -- & $2.7\pm2.1$ & multi-UV \\
CRISTAL-14 & -- & -- & -- & -- & single \\
CRISTAL-15 & -- & -- & -- & -- & single \\
CRISTAL-16 & $1.49\pm0.14$ & -- & $0.9\pm0.6$ & -- & single \\
CRISTAL-17 & -- & -- & -- & -- & single \\
CRISTAL-19 & $0.60\pm0.11$ & $1.61\pm0.94$ & $2.5\pm0.9$ & $0.9\pm0.6$ & single \\
CRISTAL-20 & $0.36\pm0.06$ & $1.50\pm0.32$ & $5.0\pm1.0$ & $1.2\pm0.3$ & single \\
CRISTAL-21 & $0.38\pm0.06$ & $7.11\pm3.06$ & $4.3\pm1.7$ & $0.2\pm0.1$ & single \\
CRISTAL-22a & $0.76\pm0.07$ & $0.88\pm0.06$ & $1.2\pm0.8$ & $1.0\pm0.7$ & pair \\
CRISTAL-22b & -- & -- & -- & -- & pair \\
CRISTAL-23a & $0.60\pm0.04$ & $3.19\pm0.61$ & $4.8\pm0.6$ & $0.9\pm0.2$ & pair \\
CRISTAL-23b & -- & -- & -- & -- & pair \\
CRISTAL-23c & -- & -- & -- & -- & single \\
CRISTAL-24 & $0.69\pm0.26$ & $1.34\pm0.18$ & $4.4\pm2.2$ & $2.2\pm0.8$ & single \\
CRISTAL-25 & $0.86\pm0.03$ & $2.10\pm0.76$ & $2.9\pm0.4$ & $1.2\pm0.5$ & single \\
\hline
  \end{tabular}
\end{table*}

\ 

\section{Results and discussion} \label{sec:Section4}
\subsection{Size comparison} \label{subsec:Section4.1}
We start by comparing the measured effective radii among different tracers (rest-frame UV and FIR continua) to characterize the spatial extent of the \cii line emission in the context of star formation activity. In Table\,\ref{table:Table2}, we list the sizes and size ratios discussed below.

\begin{figure*}
    \centering
    \includegraphics[width=0.95\linewidth]{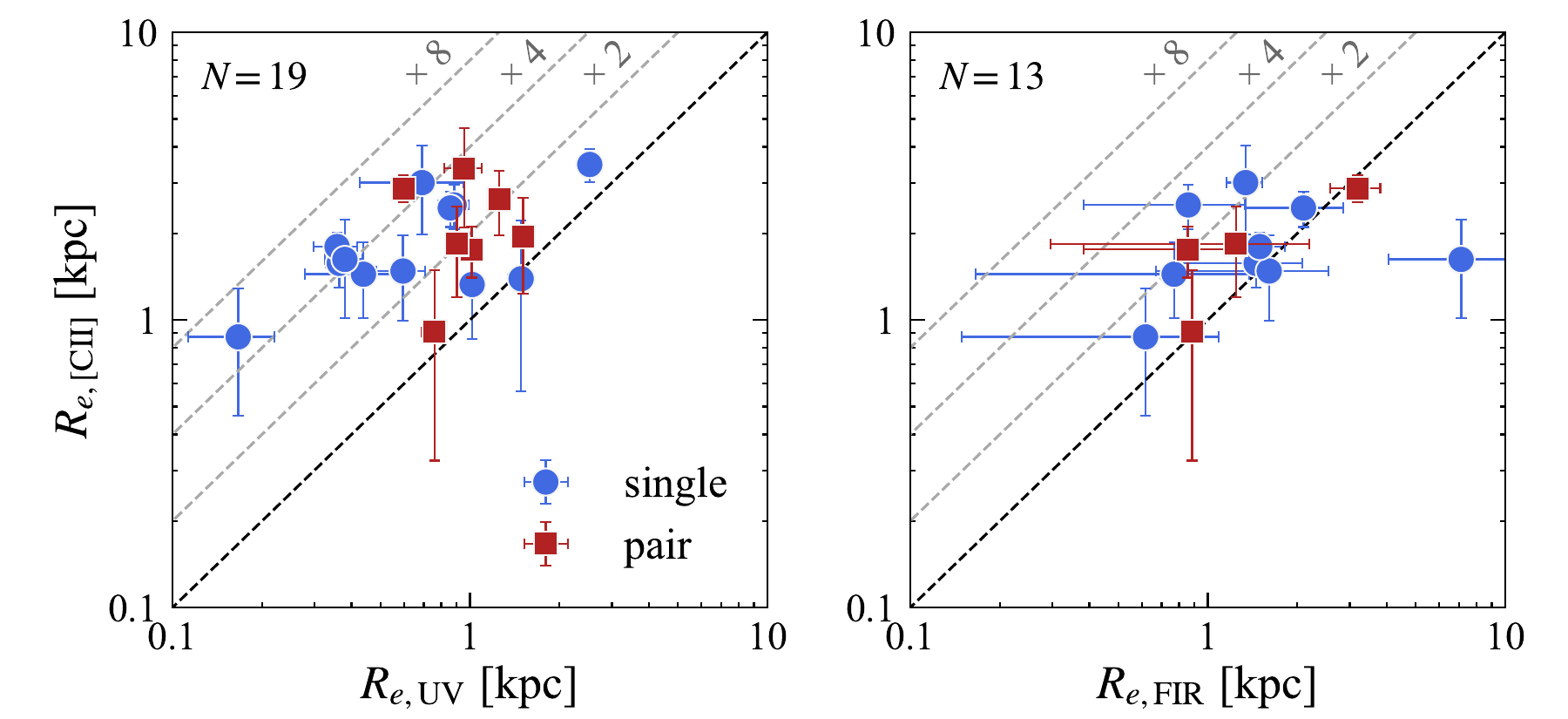}
    \caption{Comparisons of the effective radii of the \cii line and the continuum emission. (Left) Effective radius of \cii line emission compared to the rest-frame UV continuum emission (HST/WFC3 F160W). Symbols are the same as in Figure\,\ref{fig:Fig4}. The dashed black line indicates the one-to-one relation between $R_{e,[\mathrm{CII}]}$ and $R_{e,{\mathrm{UV}}}$, and three dashed gray lines above this line represent $R_{e,[\mathrm{CII}]}/R_{e,{\mathrm{UV}}}=2,\ 4,\ 8$. (Right) Same as the left panel, but \cii line radius is compared to the rest-frame FIR emission, measured from the ALMA Band-7 continuum images \citep{2024A&A...690A.197M}. {On average, the \cii sizes are larger than the rest-frame UV sizes by a factor of 2.90; compared to the rest-frame FIR sizes, they tend to be larger by a factor of 1.54 although with large uncertainties.}}
    \label{fig:Fig5}
\end{figure*}

The left panel of Figure\,\ref{fig:Fig5} shows a comparison between the \cii line and rest-frame UV continuum radii. From Figure\,\ref{fig:Fig5} (left), we find that the \cii line emission is systematically more extended than the rest-frame UV emission, which is well consistent with the previous findings (\citealp{2018MNRAS.478.1170C}; \citealp{2020ApJ...900....1F}; \citealp{2022ApJ...934..144F}). The size ratio ($R_{e,{\rm [CII]}}/R_{e,{\rm UV}}$) ranges from 0.94 (CRISTAL-16) to 5.03 (CRISTAL-20) with the average value and standard deviation of $\langle R_{e,\mathrm{[CII]}}/R_{e,{\mathrm{UV}}}\rangle=2.90\pm1.40$. We note that the median (average) positional offset between the \cii line and rest-frame UV continuum emission is 0.11\,(0.14) arcsec ($\sim0.5$\,kpc in a physical scale), which is smaller than the typical synthesized beam size of the CRISTAL data. We interpret this as them having consistent peak positions. 

We exclude multiple-UV and multiple-[C\,{\sc ii}] systems in this diagram, because each \cii line or UV component may arise from multiple galaxies for these classes, making the interpretation difficult. {Among the multiple-UV systems, the central UV component in CRISTAL-01a and CRISTAL-01b was successfully modeled by GALFIT, whereas the remaining multiple-UV systems failed to converge due to contamination from nearby UV sources. We determine [C\,{\sc ii}]-to-UV size ratios of 3.0 and 3.8 for CRISTAL-01a and CRISTAL-01b, respectively, which are both slightly larger than the average value. \cite{2017MNRAS.466.3612B} report larger UV sizes in bright UV galaxies with clumpy structures ($M_{\mathrm{UV}}\lesssim-21.5$\,mag), suggesting a similar effect may influence the \cii line emission in these multiple-UV systems (Figure\,\ref{fig:FigA1}).}

In the right panel of Figure\,\ref{fig:Fig5}, we show the size comparison between the \cii line and FIR continuum emission. The latter was measured in visibility modeling with a circular exponential disk model (\citealp{2024A&A...690A.197M}; Section\,\ref{subsec:Section3.2.2}). The best-fit central positions of the \cii line and FIR continuum emission are in good agreement, with the median (average) offset of 0.09\,(0.12) arcsec, again smaller than the beam size of the CRISTAL data. Similarly to the comparison with the UV continuum emission, we find that the \cii line emission is comparable to, or more extended than the FIR continuum emission. One galaxy that is well below the one-to-one line is CRISTAL-21, which has an unreliable FIR size measurement due to the insufficient S/N of the visibility data, particularly at short $uv$ distances. Therefore, hereafter, we do not include the results of CRISTAL-21 when the FIR radius is related to the discussions. We calculate the average size ratio and standard deviation of $\langle R_{e,\mathrm{[CII]}}/R_{e,{\mathrm{FIR}}}\rangle=1.54\pm0.63$. Within the uncertainty, this size ratio suggests that the \cii line radius is consistent with the FIR continuum radius. In fact, we find that two-thirds of the sample (eight out of 12) shown in Figure\,\ref{fig:Fig5} have comparable FIR continuum radius with \cii line radius ($R_{e,\mathrm{[CII]}}/R_{e,\mathrm{FIR}}<1.5$). 

Three recent studies have independently reported similar results. \cite{2024A&A...683L...9R} performed a stacking analysis of \cii line, FIR continuum (160\,$\mu$m), and near-UV (NUV; 2300\,\AA) continuum emission in local dwarf galaxies considered as analogs of high-redshift galaxies. They discovered that the \cii line and FIR emission is more extended than the NUV emission. The single S\'{e}rsic fittings of \cii line and FIR continuum show an excellent agreement in effective radius with a Gaussian-like profile (S\'{e}rsic index $n\sim0.5$). \cite{2024A&A...686A.187P} reported the same comparison for four galaxies drawn from the ALPINE Survey resulting in a median (average) size ratio of $R_{e,\mathrm{[CII]}}/R_{e,{\mathrm{FIR}}}=1.29\pm0.14$ ($1.24\pm0.19$), which is slightly smaller than our measurements. Two out of four galaxies are CRISTAL-24 and CRISTAL-25, and we confirm that the measurements are consistent with ours within uncertainties. Given that both the average size ratio and its standard deviation derived from the CRISTAL sample are larger than the values reported by \cite{2024A&A...686A.187P}, we are probing more diverse populations in terms of spatial relation between \cii line and FIR continuum emission, likely due to the improved spatial resolution and sensitivity of the CRISTAL data. Finally, \cite{2021A&A...654A..37D} reported an average size ratio of $\langle R_{e, \mathrm{[CII]}}/R_{e, \mathrm{FIR}}\rangle=1.61\pm0.10$ for four dust-obscured quasars at $z\simeq3.0-4.6$, which are a few orders of magnitude brighter in the IR luminosity compared to our sample (see also \citealp{2020ApJ...904..130V}). All these studies imply that the range of [C\,{\sc ii}]-to-FIR size ratios at high redshift is relatively narrow ($1\lesssim R_{e, \mathrm{[CII]}}/R_{e, \mathrm{FIR}}\lesssim2$), compared to that of [C\,{\sc ii}]-to-UV size ratios which can be affected by dust obscuration.

One possible origin of extended \cii line emission is ongoing merging activity. When diffuse carbon-rich gas is morphologically disturbed, or taken away from the stellar component, the \cii line emission can become more extended compared to the rest-frame UV continuum. Such a phenomenon has been exemplified in the case of {vuds\_cosmos\_5101209780 \citep{2020A&A...643A...7G} and CRISTAL-05 \citep{2024arXiv240303379P} as well as a larger sample of merging galaxies in the ALPINE sample \citep{2024A&A...690A.255D}. \cite{2020A&A...643A...7G} and \cite{2024A&A...690A.255D} report that $\sim$\,25--50\,\% of the total \cii line emission is coming from a diffuse gas envelope. The fraction of \cii line emission originating from the envelope is defined by subtracting the \cii line flux of the ```galaxy''' component from the total \cii line flux, characterized by spectra extracted from regions above $1\sigma$ or $2\sigma$ significance level in the moment-0 maps. These studies highlight the impact of early metal-enrichment at large radii, though the definition depends on the sensitivity of the observations. In particular, extended line flux with low $\mathrm{S/N}$ may not be CLEANed if the stopping threshold of {\tt tclean} is higher than this flux, potentially leading to an overestimation of the extended and diffuse flux.} For galaxies classified as single, we derive $\langle R_{e,\mathrm{[CII]}}/R_{e,{\mathrm{UV}}}\rangle=3.21\pm1.42$ and $\langle R_{e,\mathrm{[CII]}}/R_{e,{\mathrm{FIR}}}\rangle=1.62\pm0.69$, while for pair systems we find $\langle R_{e,\mathrm{[CII]}}/R_{e,{\mathrm{UV}}}\rangle=2.38\pm1.22$ and $\langle R_{e,\mathrm{[CII]}}/R_{e,{\mathrm{FIR}}}\rangle=1.36\pm0.45$. Given the large uncertainties of these size ratios, we refrain from claiming an ongoing physical effect from galaxy interactions. Thus, we tentatively conclude that we do not find differences in size ratios between isolated galaxies classified as single and mergers classified as pair systems. 

Since there is a tight correlation between \cii luminosity and SFR, \cii line emission has been used as an indicator of star formation activity. However, the size differences between \cii line and continuum (rest-frame UV and FIR) emission suggests that the \cii line emission is not completely spatially coupled to star formation activity, which might be at odds with studies of nearby galaxies (\citealp{2014A&A...570A.121P}; \citealp{2015ApJ...798...24K}; \citealp{2015ApJ...800....1H}). Given the average size ratios of 2.9 and 1.5, we infer that the \cii line emission is approximately two times more extended than the star formation activity traced by continuum emission. We shall further discuss this point in Section\,\ref{subsec:Section4.3}.

\begin{figure*}[ht!]
    \centering
	\includegraphics[width=\linewidth]{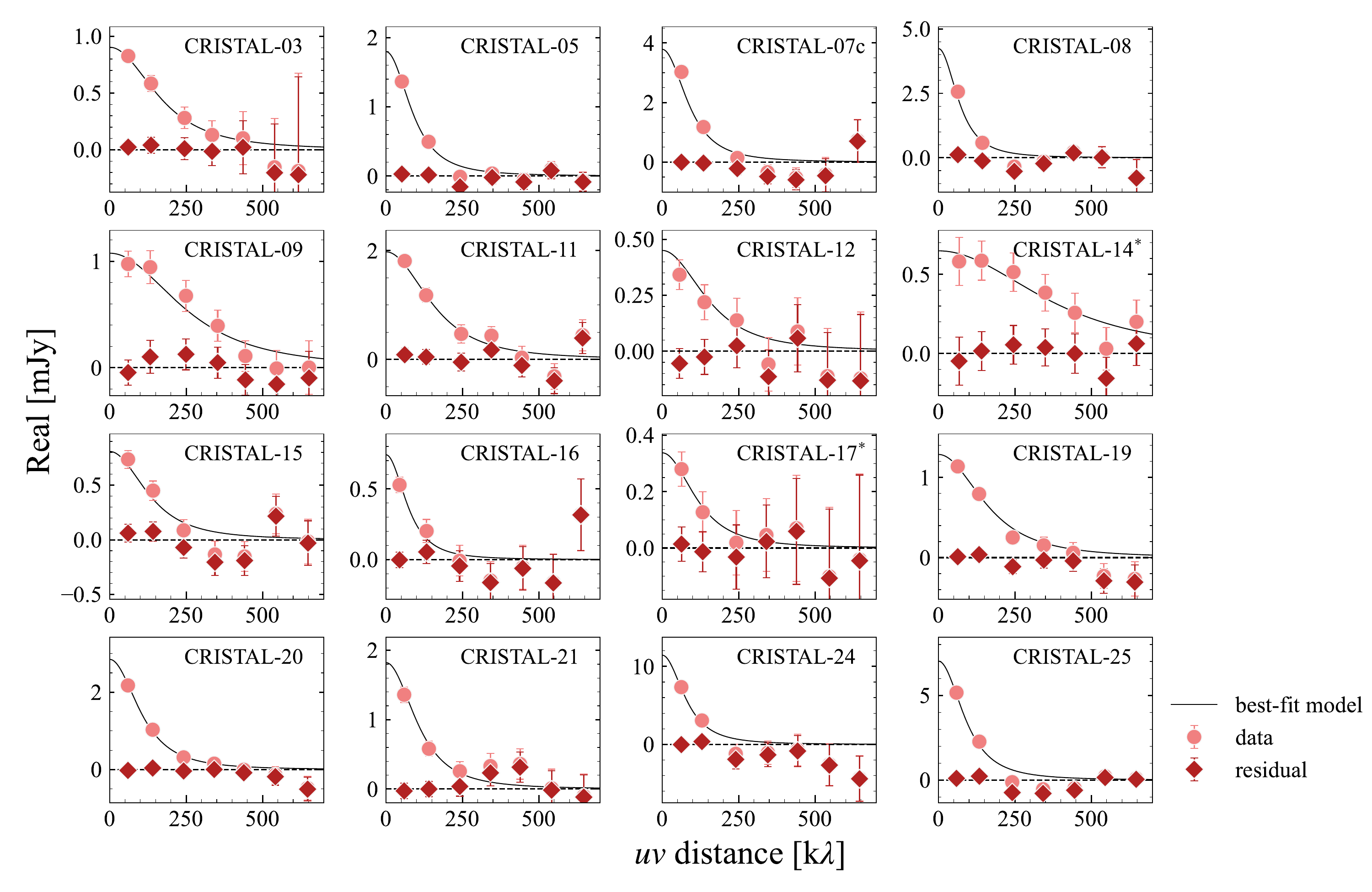}
    \caption{Flux density in the form of the real part of the visibility, as a function of $uv$ distances along the minor axis. Each data point represents the mean value per 100\,k$\lambda$ bin. The galaxies that were fit in a circular exponential disk model are marked in an asterisk. The uncertainties represent the standard deviation. Both the data (circle) and the residual (diamond) visibilities are shown. The residuals are created by subtracting the best-fit model (solid black line) from the data.}
    \label{fig:Fig6}
\end{figure*}

\begin{figure*}
    \centering
	\includegraphics[width=0.88\linewidth]{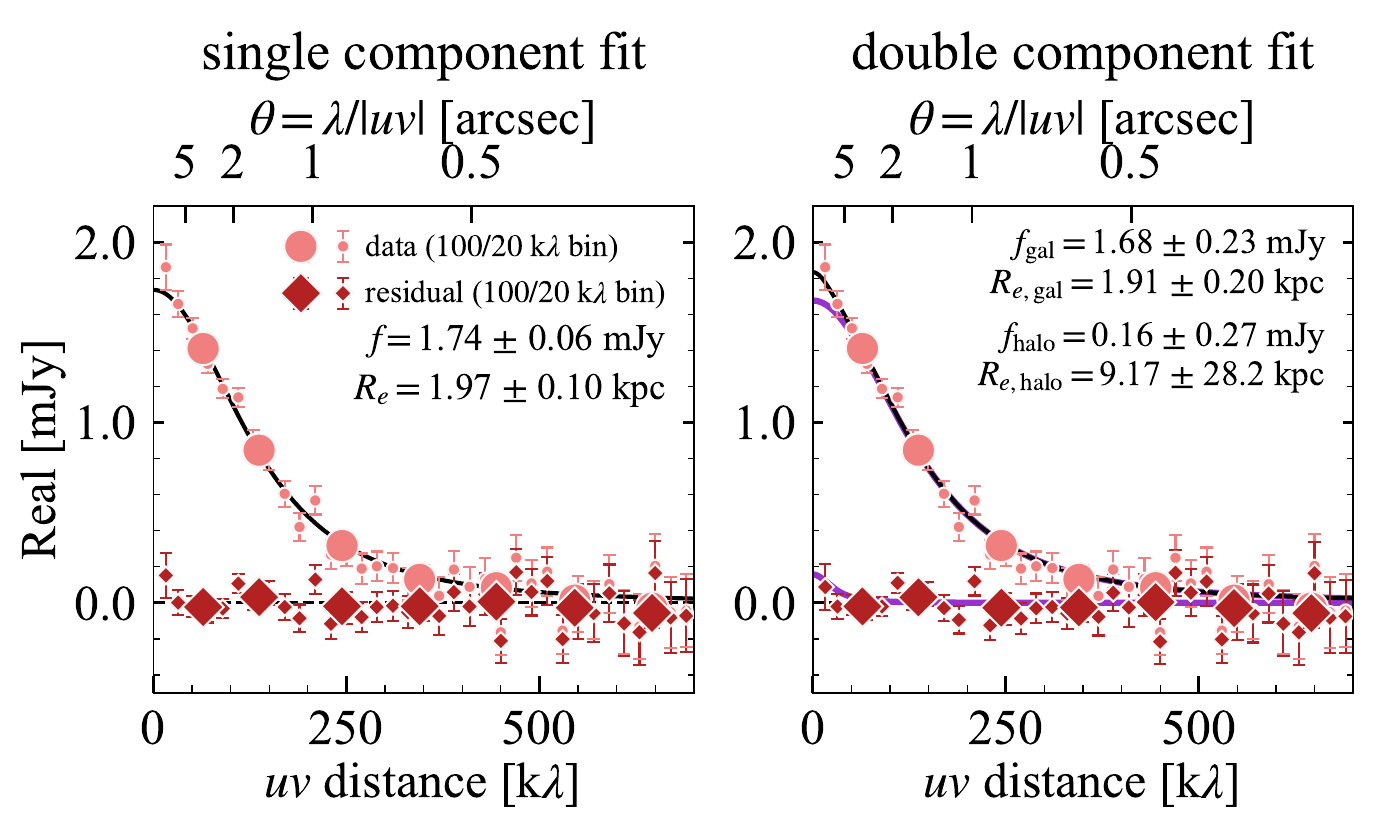}
    \caption{{{Real part of the stacked visibility (circles) and the residual visibility (diamonds) as a function of $uv$ distances for single component (left) and double component (``galaxy'' + ``halo''; right) fittings. Average values of two types of binning (20 k$\lambda$ and 100 k$\lambda$) are shown. Uncertainties represent a standard deviation of binned visibility data. The best-fit models are shown as a solid black line (total) and solid purple lines (each component). The flux and effective radius of the best-fit models are summarized at the upper right corners. At the smallest 20\,k$\lambda$ bin, the residual show an excess at $1.2\sigma$ and $0.7\sigma$ significance levels, which corresponds to 8.8\,\% and 5.5\,\% of the total flux density for single and double component fittings, respectively.}}}
    \label{fig:Fig7}
\end{figure*}

\begin{figure}[t!]
    \centering
	\includegraphics[width=0.93\columnwidth]{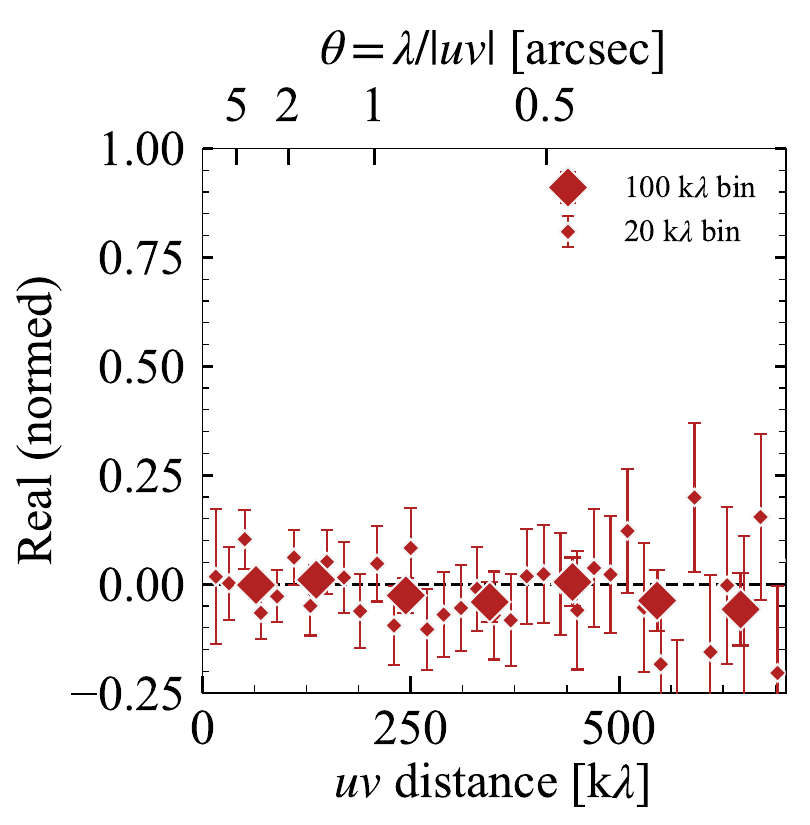}
    \caption{Real part of the stacked residual visibility as a function of $uv$ distances. Average values of two types of binning (20 k$\lambda$ and 100 k$\lambda$) are shown. Uncertainties represent a standard deviation of binned visibility data.}
    \label{fig:Fig8}
\end{figure}
\subsection{Constituent of \cii line emission} \label{subsec:Section4.2}

In this section, we assess whether \cii line emission in the SFGs has a halo structure as a secondary component. If such a halo structure is confirmed, the physical origin would be different from the central \cii component associated with star-forming ISM (e.g., star-formation-driven outflow, cold stream). Therefore, the identification of \cii halo gives us an important insight into the origin of the extended \cii line emission.

Because \cii halo is expected to have low-surface density at large radii, most of the previous studies needed to rely on the stacking analysis to confirm such structure (\citealp{2019ApJ...887..107F}; \citealp{2020A&A...633A..90G}; \citealp{2020ApJ...904..131N}). To date, the only case in an individual \cii halo has been confirmed {as an extended secondary component structure} is A1689-zD1, a strongly lensed galaxy at $z=7.13$ \citep{2022ApJ...934...64A}.
As observations of the CRISTAL Survey are deeper than the previous observations, we can now assess whether such structure can be confirmed individually. To eliminate the possibility that \cii line emission from satellite galaxies contaminates the halo component, we focus on 16 single galaxies, which we classified in Section\,\ref{subsec:Section3.1}. We do not include CRISTAL-23c since the galaxy might be in a phase of triple-merger \citep{2020MNRAS.491L..18J}.

{Figure\,\ref{fig:Fig6} visualizes the data, best-fit model, and residual as a form of visibility data for 16 single objects that we created in Section\,\ref{subsec:Section3.2}. The average velocity range of 16 single galaxies is 485.0 km/s.} The data is extracted by the {\tt getdata} function of a casatool {\tt ms}. In Figure\,\ref{fig:Fig6}, average values and standard deviations of the real part of visibility data per 100\,k$\lambda$ bin are shown. The residual data is created by subtracting the best-fit model from the data. For all galaxies, the data and the best-fit model show a good agreement, and we do not find an excess at the residual data in the shortest $uv$ distance bin ($\sim50$\,k$\lambda$). This indicates that the visibility profile can be well fit by an exponential disk model and an extended secondary component is not necessary to explain the data.  

To further confirm this, we stack the visibility data of 16 single galaxies and again perform the visibility modeling in the same manner. {We perform both single component and double component fittings. The result of the single component fitting is shown in the left panel of Figure\,\ref{fig:Fig7}.} As well as the visibility data of the individual galaxies, the stacked visibility data is fit well by a single exponential disk profile. The model is characterized by an effective radius of $R_{e,\mathrm{[CII]}}=1.97\pm0.10$\,kpc, assuming an average redshift of $z=5.22$, which is consistent with the average radius ($\langle R_{e,\mathrm{[CII]}}\rangle=1.90$\,kpc) of the entire CRISTAL galaxies. At the smallest 20\,k$\lambda$ bin of the residual visibility, we find a flux excess of 0.15\,mJy ($1.2\sigma$ significance level). {The result of the double component fitting is shown in the right panel of Figure\,\ref{fig:Fig7}. Here, we assume two concentric exponential disk components. The brighter component ($1.68\pm0.23$\,mJy; hereafter ``galaxy'' component) has a radius of $R_{e,\mathrm{gal}}=1.91\pm0.20$\,kpc, which is similar to the best-fit single component. The other component ($0.16\pm0.27$\,mJy; hereafter ``halo'' component) has an extended radius of $R_{e,\mathrm{halo}}=9.17\pm28.2$\,kpc. Since the fittings converge to the same reduced $\chi^{2}$ value, we cannot statistically favor either the single or double component model. However, the existence of an extended, diffuse component remains uncertain due to its faintness and large uncertainties. The flux of the ``halo'' component constitutes $8.7\pm13.4$\,\% of the total flux, and this is smaller than what has been suggested in previous \cii line observations (e.g., $\sim30$\,\% for A1689-zD1 reported in \citealp{2022ApJ...934...64A}, but see \citealp{2023MNRAS.518..691J} for faint gaseous halo detected in CO $J=3-2$ line emission). The flux becomes negative within the range of the uncertainty, and the standard deviation of $\pm\,0.13$\,mJy at the smallest 20\,k$\lambda$ bin gives the detection of $1.2\sigma$ level, which both point out that the detection is not significant. As we have identified various axis ratios ($q=0.31-0.90$) in our sample (Table\,\ref{table:Table1}), it is possible that the extended \cii line emission along the major axis is diluted by the stacking and create the ``halo'' component artificially.}

{Alternatively, we now focus on the stacked residual of the visibility data where the main disk component obtained from a single exponential disk fitting is subtracted. 
Figure\,\ref{fig:Fig8} shows the stacked residual visibility as a function of $uv$-distances. In order to avoid the stacked visibility data being skewed by the bright objects in which the amplitude of residual visibility is also bright, we normalize the residual visibility by the total \cii flux obtained from the best-fit model before the stacking. We show the average values per both $20$\,k$\lambda$ and $100$\,k$\lambda$ bins, but no clear excess can be found in either case. By adopting standard deviations of visibility data within each bin as uncertainties, the small deviations from the horizontal axis at shorter-$uv$ distances ($\lesssim200$\,k$\lambda$) are below the $1\sigma$ level. If the fitting results are dragged by data from the shorter $uv$ distances, we would find an excess at larger $uv$-distances as a remnant of the fitting, but no such feature can be found. 

{To summarize, from the double component fitting in the stacked visibility data (Figure\,\ref{fig:Fig7} right), we do not find a statistically significant evidence of \cii halo as a secondary component.
The modeled secondary component consists of $8.7\pm13.4$\,\% of the total flux, which is smaller by several times than what has been suggested in the previous \cii line observations at high redshift.  Based on Figure\,\ref{fig:Fig6}, \ref{fig:Fig7}, and \ref{fig:Fig8}, 
we conclude that the \cii line emission of the CRISTAL galaxies can be adequately explained by a single exponential disk component.}}

\subsection{What determines the spatial extent of the \cii line?} \label{subsec:Section4.3}

{In this section, we discuss the origin of extended \cii line emission. The correlation between \cii luminosity and SFR, which holds true for several orders of magnitude, suggests that the \cii line emission can be considered a good tracer of PDRs on molecular cloud surface ($n\simeq10^{3}$--$10^{5}$\,cm$^{-3}$, $T\simeq10^{2}$\,K) surrounding young massive stars. However, \cii line emission can arise from more diffuse gaseous components ($n\lesssim10^{3}$\,cm$^{-3}$) than PDRs. Following the description of the ISM conditions in \cite{1997ApJ...483..200M}, these are cold neutral medium ($n\simeq10^{2}$\,cm$^{-3}$, $T\simeq10^{2}$\,K), warm neutral medium ($n\simeq1$\,cm$^{-3}$, $T\simeq5\times10^{3}$\,K), and ionized medium ($n\lesssim1$\,cm$^{-3}$, $T\simeq10^{4}$\,K; Section\,\ref{subsubsec:Section4.3.2}). In neutral ISM conditions, atomic hydrogen would be the dominant collision partner, while in PDRs either atomic or molecular hydrogen can be effective in the \cii excitation \citep{2012ApJS..203...13G}.}

{As FIR continuum emission traces a dust re-emission, the spatial extent of FIR continuum emission can be regarded as regions where PDRs are dominant. Hence, the comparable spatial extent between the \cii line and FIR continuum emission ($R_{e,{\mathrm{[CII]}}}/R_{e,{\mathrm{FIR}}}\approx1$) indicates that the \cii line emission coincides with star formation activity traced by the PDRs.}
\cite{2021ApJ...918...69U} report the \cii line and FIR continuum observations of galaxies in the SSA22 protocluster at $z=3.1$ and discuss that \cii line emission is mainly associated with star formation activity, although other origins and potential effects from the overdense environment are not ruled out. \cite{2020ApJ...904..131N} also draw a similar conclusion for $z>6$ quasars. However, as those samples are limited to IR-bright galaxies with $L_{\mathrm {IR}}\gtrsim10^{12} L_{\odot}$, the results may not apply to typical SFGs ($L_{\mathrm {IR}}\lesssim10^{12} L_{\odot}$) at corresponding redshifts. Previous studies have reported few galaxies showing $R_{e,{\mathrm{[CII]}}}/R_{e,{\mathrm{FIR}}}\approx1$ (\citealp{2020ApJ...900....1F}; \citealp{2024A&A...686A.187P}), and thanks to the sensitivity of ALMA observations taken by the CRISTAL Survey, we can now postulate whether such galaxies comprise of the majority of the SFGs at $z\sim4-6$, which is consistent with the general picture that the \cii line emission is coming from the PDRs.

From the size comparisons that we presented in Section\,\ref{subsec:Section4.1}, we find that the sizes of the \cii line emission in our sample are on average 2.90 and 1.54 times larger compared to the rest-frame UV and FIR continuum emission, respectively (Figure\,\ref{fig:Fig5}), suggesting that \cii line emission is approximately twice more extended than the main star-forming regions. This indicates that the contribution from the PDRs is not sufficient to explain the \cii line emission beyond the radius of star-forming regions. 
For simplicity, if we assume that the \cii radius is exactly twice more larger than the radius that encloses half of the SFR, and the contribution from PDRs is dominant (100\,\%) within the \cii radius ($r<R_{e,\mathrm{[CII]}}$), the average contribution from PDRs at $r>R_{e,\mathrm{[CII]}}$ is roughly estimated to be $\approx20$\,\%. When spatially integrated, the PDR fraction is then $\approx60$\,\%. Low contribution from PDRs at the outskirts of the galaxies is qualitatively consistent with \cite{2024ApJ...976...70L}, who find that {the spatially resolved $\Sigma_{\mathrm{[CII]}}-\Sigma_{\mathrm{SFR}}$ has a steeper slope in about half of the CRISTAL galaxies that are examined than the ones found in nearby star-forming galaxies (e.g., \citealp{2014A&A...568A..62D}; \citealp{2015ApJ...800....1H}). While this could be caused by the [C\,{\sc ii}] deficit in the central regions, since we do not find such evidence as we show later (see Figure\,\ref{fig:Fig10} in Section\,\ref{subsubsec:Section4.3.3}; see also \citealp{2021A&A...649A..31H} and \citealp{2024arXiv240303379P} for individual cases), we suggest that the $\Sigma_{\mathrm{[CII]}}-\Sigma_{\mathrm{SFR}}$ does not hold at large radii as the central region tends to deviate toward brighter \cii surface densities.}

In the Milky Way, \cite{2013A&A...554A.103P} report that about 20\,\% of the \cii line emission arises from the cold neutral medium. In M\,33, \cite{2013A&A...553A.114K} discover an increasing contributions from a cold neutral medium to the \cii line emission with radius. For Haro\,11, a nearby dwarf galaxy that has similar \cii and IR luminosities to the CRISTAL galaxies, \cite{2012A&A...548A..20C} report a contribution from PDRs to the \cii line emission of 10\,\%, while the rest of the \cii line emission is coming from diffuse low-ionization gas. \cite{2016AJ....152...51D} compare the spectra of [C\,{\sc ii}], CO\,$J$\,=\,1--0, and H\,{\sc i} lines of a sample of 10 nearby galaxies. They find that the radial profile of the \cii line emission is taking an intermediate slope between CO\,$J$\,=\,1--0 and H\,{\sc i} lines (see also \citealp{2021ApJ...915...92T}). These studies motivate us to think that by measuring the spatial extent of \cii line emission at a kiloparsec-scale resolution, we are looking at the intermediate spatial extent between molecular and atomic gas distributions. Considering that the average H\,{\sc i} gas fraction ($M_{\mathrm{HI}}/(M_{\mathrm{HI}}+M_{\star})$) monotonically increases with redshift, from 25\,\% at $z=0$ to 70\,\% at $z\sim$\,4--6 (\citealp{2020ApJ...902..111W}; \citealp{2021ApJ...922..147H}), and that H\,{\sc i} gas is thought to be extended up to a circumgalactic scale along the filaments of the cosmic web (e.g., \citealp{2017MNRAS.468.4831K}), the extended \cii line emission in the CRISTAL galaxies is likely to be a combination of that coming from PDRs and cold/warm neutral medium, where most of the mass can be traced by molecular and atomic hydrogen, respectively. {In nearby galaxies, the dust mass surface density correlates with the total gas mass surface density better than the molecular or atomic surface densities at a kiloparsec scale (e.g., \citealp{2022A&A...668A.130C}). Since the atomic gas distribution is generally more extended than that of the molecular gas, the strong correlation between dust and total gas mass suggests that at large radii, dust can be a good tracer of atomic gas rather than molecular gas. {While we cannot assume the same correlation to the CRISTAL galaxies due to the differences in gas fraction and metallicity, this correlation in nearby galaxies supplements our argument that the contribution from atomic gas to the \cii line emission becomes increasingly dominant with radius.}

Numerous theoretical works have been presented to constrain the properties of FIR fine-structure lines at high redshift (e.g., \citealp{2006ApJ...647...60N}; \citealp{2013MNRAS.433.1567V, 2015ApJ...813...36V}; \citealp{2017MNRAS.468.4831K}; \citealp{2017MNRAS.471.4128P}; \citealp{2018ApJ...857..148O}; \citealp{2018A&A...609A.130L}; \citealp{2019MNRAS.482.4906P}; \citealp{2020MNRAS.498.5541A}; \citealp{2023A&A...679A.131R}; \citealp{2024ApJ...965..179G}; \citealp{2024arXiv241014284C, 2024A&A...689A.106C}; \citealp{2024A&A...690A.392M}). Some of the studies that align with the discussion above include \cite{2023A&A...679A.131R} who simulated the dominant fraction (up to $\approx$70--90\,\%) of \cii line emission coming from the atomic gas phase in high-redshift galaxies (see also \citealp{2024arXiv241014284C}). Similarly, \cite{2024ApJ...965..179G} report that the contribution from warm neutral medium (not necessarily molecular gas) is dominant ($\approx30-40$\,\%) particularly at sub-solar metallicity while H\,{\sc ii} region may be comparably important, based on high-resolution hydrodynamical simulations of a self-regulated ISM. In contrast, \cite{2015ApJ...813...36V} discuss that the \cii line emission from cold gas may be attenuated by cosmic microwave background (CMB) by a factor of 0.1--0.2, because the CMB temperature approaches the spin temperature of the \cii line transition in the cold neutral medium ($\sim20$\,K). This results in a minor contribution of cold neutral gas ($\lesssim10$\,\%) to the \cii line emission (see also \citealp{2018ApJ...857..148O}; \citealp{2018A&A...609A.130L}). It should be noted that these simulated galaxies are affected by slightly higher CMB temperatures since they are at higher redshifts ($z\gtrsim6$) than our sample.

\begin{figure*}
    \centering
	\includegraphics[width=0.92\linewidth]{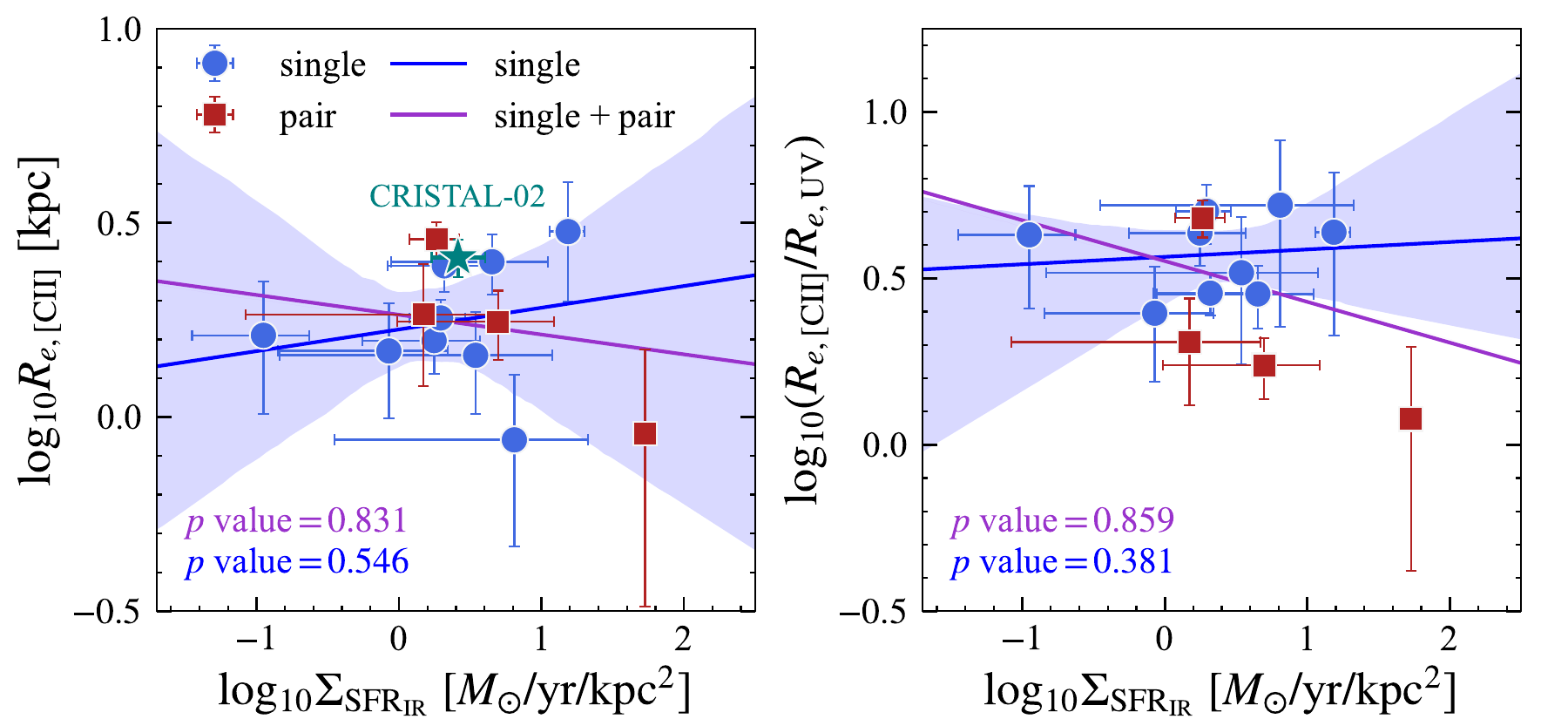}
    \caption{{Dependence of the spatial extent of \cii line emission on the $\mathrm{SFR_{IR}}$ surface density. (Left) \cii line radius as a function of the $\mathrm{SFR_{IR}}$ surface density. The solid blue and purple lines show a best-fit linear regression for single and single$+$pair sources, respectively. The shaded region shows a 95\,\% confidence level of the fit for single galaxies. The $p$\,values of a Spearman's rank correlation test for both single and single$+$pair parent samples are shown in the lower-left corner. We also highlight CRISTAL-02 (multiple-UV), a galaxy which shows an evidence of broad \cii wing component in multiple spatially resolved elements of the galaxy (Davies et al. in prep). (Right) Same as the left panel, but taking the [C\,{\sc ii}]-to-UV size ratio as the vertical axis. Given the $p$\,values, we cannot rule out the null hypothesis that there is no correlation between these parameters.}}
    \label{fig:Fig9}
\end{figure*}

The cause of variations in $R_{e,\mathrm{[CII]}}/R_{e,\mathrm{FIR}}$ among individual galaxies is still not clear. If extended \cii line emission originates from a diffuse neutral medium, then it is possible that the \cii line emission can be taken away from the host galaxies during merger events, as has been demonstrated in simulated galaxies \citep{2022MNRAS.513.5621P}. While we fail to statistically confirm that pair systems have more extended \cii sizes than others (Section\,\ref{subsec:Section3.2}), JWST images reveal that some single galaxies identified by HST images are actually composed of multiple stellar components \citep{2024arXiv240910963L}. This includes two galaxies, CRISTAL-07c and CRISTAL-11, which have $R_{e,\mathrm{[CII]}}/R_{e,\mathrm{FIR}}>2$. Future assessments of how the $R_{e,\mathrm{[CII]}}/R_{e,\mathrm{FIR}}$ (and $R_{e,\mathrm{[CII]}}/R_{e,\mathrm{UV}}$) varies during past merger events using zoom-in cosmological simulations will give an insight into our results. {Alternatively, the variations in $R_{e,\mathrm{[CII]}}/R_{e,\mathrm{FIR}}$ in our sample may reflect the presence of extraplanar gas and dust, as is demonstrated by the observations of NGC\,891, a nearby edge-on galaxy. \cite{2020ApJ...902...28R} report that the \cii line emission in a thick disk of NGC\,891 has larger scale height (2.8\,kpc) compared to mid-IR emission tracing PAHs, as well as FIR emission (1.4\,kpc; \citealp{2016A&A...586A...8B}). Nonetheless, the data with deeper sensitivity and higher spatial resolution are required to verify this scenario in our sample.}

In the following, we discuss four alternative mechanisms that could be the cause of extended \cii line emission in addition to diffuse neutral gas: outflows (Section\,\ref{subsubsec:Section4.3.1}), ionized gas (Section\,\ref{subsubsec:Section4.3.2}), radial variation in the \cii deficit (Section\,\ref{subsubsec:Section4.3.3}), and shock heating (Section\,\ref{subsubsec:Section4.3.4}).

\subsubsection{Outflows} 
\label{subsubsec:Section4.3.1}

One of the supporting scenarios for the extended \cii line emission is ongoing outflows driven by star formation activity. Semi-analytical models predict that cold-mode outflows could reproduce the extended \cii line emission \citep{2020MNRAS.495..160P,2023MNRAS.519.4608P}, and this has been observationally supported by the spectra stacking analysis of \cii line emission for {both star-forming galaxies at $4<z<6$ (\citealp{2018MNRAS.473.1909G}; \citealp{2020A&A...633A..90G})} and local low-metallicity dwarf galaxies \citep{2023A&A...677A..44R}. On the other hand, \cite{2022ApJ...934...64A} do not find an outflow signature in \cii line from a strongly lensed galaxy A1689-zD1 at $z=7.13$, but detect a broad component in \oiii line emission, indicating the presence of ionized outflowing gas that will subsequently cool and emit the \cii line emission at larger radius (see also \citealp{2022ApJ...929..161W}). For $z>6$ quasars, \cite{2020ApJ...904..131N} report null detection of broad spectral feature from \cii line based on spectral stacking (see also \citealp{2018ApJ...854...97D}).

As such, it is still not clear whether there is a physical association between outflows and the extended \cii line emission seen in projected sky distribution at $z>4$. The situation has become more puzzling, since recent ALMA observations (\citealp{2020ApJ...905...85S}; \citealp{2024ApJ...962....1S}; \citealp{2024ApJ...963...19N}) report the detection of a blueshifted OH\,119$\mu$m absorption feature as evidence of molecular outflows, while a broad \cii line feature is not detected for the same sources. These results indicate that the \cii line emission is not always a robust tracer of molecular outflows. \cite{2023ApJ...958..109L} present a \cii line observation of the outflow region of a nearby starburst galaxy M\,82. The authors find that the fraction of \cii line flux associated with the outflow region is about one-quarter of the total \cii line emission, and the majority ($>55$\,\%) of the \cii line flux in the outflow region arises from the atomic component. If the composition of the \cii line in high-redshift galaxies is similar to M\,82, it is not surprising to have a weak or no detection of the broad \cii line feature.

It has also been suggested that AGN outflows can enhance \cii line emission due to an additional contribution from either X-ray-dominated or ionized regions (\citealp{2019A&A...626L...3S}, but also see \citealp{2010ApJ...724..957S}). Since the original sample selection in the ALPINE Survey excluded Type\,I AGN with broad optical spectral line \citep{2020A&A...643A...1L}, our sample similarly does not include bright Type\,I AGN. However, \cite{2024ApJ...976...70L} discuss that CRISTAL-25 may be an AGN, because the spatially resolved SED fitting analysis shows poor fit at the central region, possibly due to the contamination of broad emission lines from AGN to the broad band photometry. Nonetheless, we confirm that \cii line emission in CRISTAL-25 is not particularly bright compared to the FIR continuum ($L_{\mathrm{[CII]}}/L_{\mathrm{FIR}}=3.7\times10^{-3}$), and so the possible effect of AGN on the enhancement of \cii line emission is not likely playing a role in this galaxy.

In this study, we are likely measuring the sizes of \cii line emission arising from both galactic ISM and outflows, if they exist. The median velocity width we use to measure the spatial extent of the \cii line emission is 440\,km/s, which is broad enough that star-formation-driven outflows can contribute to the observed structure of the \cii line emission. \cite{2019ApJ...886...29S} report a positive correlation between SFR and outflow velocity up to $z\sim6$. A similar correlation between SFR {(or its surface density $\Sigma_{\mathrm{SFR}}$)} and mass outflow rate is also reported for nearby galaxies (e.g., \citealp{2019MNRAS.483.4586F}; see also \citealp{2020A&ARv..28....2V}; \citealp{2020ARA&A..58..661F} and references therein).

Among the CRISTAL galaxies, CRISTAL-02 only shows a strong broad wing component in the \cii line spectrum (Davies et al. in prep) and weak evidence of broad wing component is detected when the \cii spectra are stacked among the CRISTAL galaxies (Birkin et al. in prep). Based on these findings, if outflows contribute to the extended \cii line emission, we might see a correlation between \cii sizes and galactic properties.

Figure\,\ref{fig:Fig9} shows the \cii size and [C\,{\sc ii}]-to-UV size ratio as a function of the $\mathrm{SFR_{IR}}$ surface density ($\Sigma_{\mathrm{SFR_{IR}}}=\mathrm{SFR_{IR}}/2\pi R_{e,\mathrm{FIR}}^{2}$) of the CRISTAL galaxies. We use the values of $\mathrm{SFR_{IR}}$ measured in \cite{2024A&A...690A.197M} who convert FIR continuum flux ($S_{158\,\mu\mathrm{m}}$) into IR luminosity ($L_{\mathrm{IR}}$) and then into obscured SFR ($\mathrm{SFR_{IR}}$) following the same method as \cite{2020A&A...643A...2B}. According to a linear regression fitting, we find a positive trend for single galaxies in both panels in Figure\,\ref{fig:Fig9}, but a negative trend when pair galaxies are included. Nonetheless, the Spearman's rank correlation tests give the $p$\,values of 0.546 (single fit) and 0.831 (single$+$pair fit) for $\log_{10}{\Sigma_{\mathrm{SFR_{IR}}}}-\log_{10}{R_{e,\mathrm{[CII]}}}$ relation (Figure\,\ref{fig:Fig9} left), and 0.381 (single fit) and 0.859 (single$+$pair fit) for $\log_{10}{\Sigma_{\mathrm{SFR_{IR}}}}-\log_{10}{(R_{e,\mathrm{[CII]}}/R_{e,\mathrm{UV}})}$ relation (Figure\,\ref{fig:Fig9} right). We similarly investigated the \cii size and [C\,{\sc ii}]-to-UV size ratio as a function of SFR, but none of the correlations were significant ($p$\,value\,$>0.05$). Therefore, regardless of the merging events, we fail to rule out the null hypothesis that there is no association between star-formation-driven (atomic) outflows and the extended \cii line emission. In addition, we find that CRISTAL-02 does not have particularly large \cii size nor $\mathrm{SFR_{IR}}$ surface density among the CRISTAL galaxies. These results may imply that there is no strong association between (atomic) outflows and the extended \cii line emission.

\subsubsection{Contribution from ionized gas} 
\label{subsubsec:Section4.3.2}

Due to the low-ionization potential of carbon, \cii line could originate from ionized regions. Because nitrogen has a higher ionization potential than hydrogen, the \nii line, which has a similar critical density as the \cii line when collisionally excited by electrons, has been widely used to discern the contribution of the \cii line from ionized gas. {Alternatively, \oiii can also be used as a tracer of ionized gas (e.g., \citealp{2018ApJ...861...94H}; \citealp{2020ApJ...896...93H}), but unfortunately the \oiii line from the redshifts of our sample is hardly accessible through the atmospheric windows.}

\begin{figure*}[ht!]
    \centering
	\includegraphics[width=0.95\linewidth]{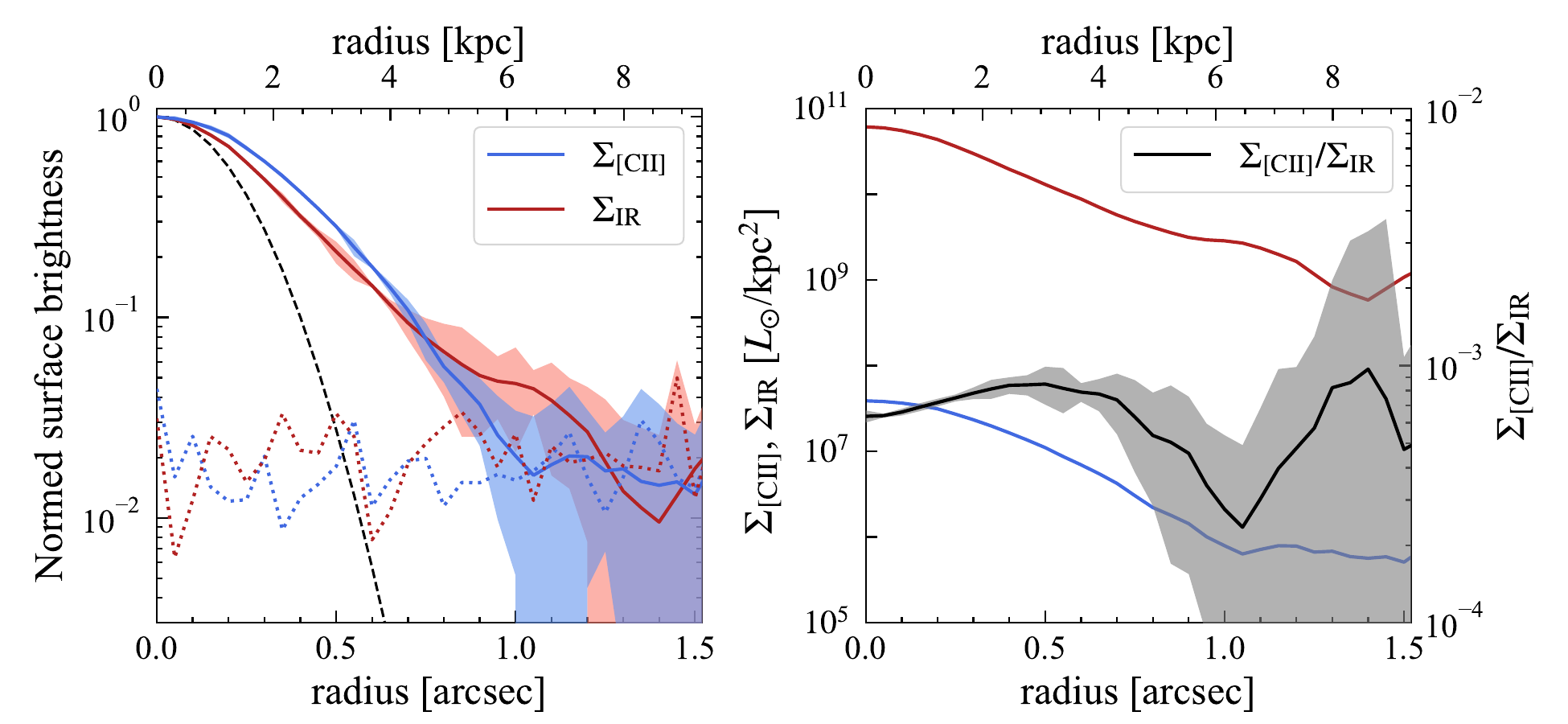}
    \caption{Radial profiles of the stacked \cii line and FIR continuum emission. (Left) {Normalized radial surface density profile of the $uv$-stacked image in \cii line ($R_{e,\mathrm{[CII]}}=1.97\pm0.17$\,kpc) and FIR continuum ($R_{e,\mathrm{FIR}}=1.53\pm0.22$\,kpc) emission.} The dotted lines show average surface density derived by setting random apertures on blank regions, which are also shown in color-shaded regions as uncertainties. The dashed black curve represents the clean PSF. The top axis is converted to a physical scale assuming an average redshift of $z=5.22$ among the stacked sample.  (Right) Absolute radial surface density profiles of the \cii line and FIR continuum emission, in which the unit is converted to $L_{\odot}/{\mathrm{kpc^{2}}}$. The solid black line shows the ratio between the \cii line and IR emission (axis shown on the right hand side). {As we do not see a strong radial variation in the [C\,{\sc ii}]-to-IR luminosity ratio, we suggest that the \cii deficit is not responsible for the extended \cii line emission.}}
    \label{fig:Fig10}
\end{figure*}

Studies of nearby galaxies with measurements of both \cii and \nii lines suggest that ionized gas, either diffuse ionized gas or H\,{\sc ii} regions, is not the main contributor of \cii line emission (e.g., \citealp{2017ApJ...845...96C}; \citealp{2018ApJ...861...94H}; \citealp{2019ApJ...886...60S}; \citealp{2021ApJ...915...92T}). Nearby luminous infrared galaxies (LIRGs) are reported to have a PDR fraction of \cii line larger than 60\,\%, up to 95\,\% \citep{2017ApJ...846...32D}. For high-redshift galaxies, \cite{2016ApJ...832..151P} study \nii line for two Lyman-break galaxies, CRISTAL-02 (LBG-1) and CRISTAL-22 (HZ10), and report higher [C\,{\sc ii}]-to-[N\,{\sc ii}] luminosity ratio for CRISTAL-02, which is less dusty than CRISTAL-22 (see also \citealp{2017ApJ...834L..16U}; \citealp{2020MNRAS.494.4090C}). The estimated contributions from ionized gas are $\sim5$\,\% and 10--25\,\% for CRISTAL-02 and CRISTAL-22, respectively. 

\cite{2017ApJ...845...96C} find a trend of decreasing [C\,{\sc ii}]-to-[N\,{\sc ii}] luminosity ratio with decreasing gas-phase metallicity in the local SFGs; that is, a low contribution to the \cii line emission from the ionized gas at high metallicity (see also \citealp{2019A&A...626A..23C}). Assuming that the mass-metallicity relations \citep{2023ApJS..269...33N} hold true in the CRISTAL galaxies, the empirical relation presented in \cite{2017ApJ...845...96C} provides an ionized fraction of 13\,--27\,\% in a stellar mass range of $\log(M_{\star}/M_{\odot})=[9.5,11.0]$. This agrees with the estimated ionized fraction of CRISTAL-22 with $\log(M_{\star}/M_{\odot})=10.35\pm0.37$ \citep{2024A&A...690A.197M}. Since the majority of the CRISTAL galaxies have lower stellar masses than this, we infer that the \cii line flux contributed from the ionized medium is not as significant as $\sim25$\,\%, indicating that the contribution from ionized gas has little impact on the spatial extent of \cii line emission.

However, some theoretical studies show mixed results regarding the impact of diffuse ionized gas on \cii line emission (e.g., \citealp{2018ApJ...857..148O}; \citealp{2023A&A...679A.131R}). Future high-resolution \nii observations will be crucial to elucidate the spatial distribution of \cii line emission associated with ionized gas at high redshift. 

\subsubsection{\cii deficit in the central region of galaxies} 
\label{subsubsec:Section4.3.3}

It is known that the [C\,{\sc ii}]-to-(F)IR luminosity ratio varies depending on the ISM conditions; for instance the ratio is particularly low in regions where (F)IR surface density is high. Several physical explanations have been made, such as low photoelectric heating efficiency due to small grain destruction or charging, high ionization parameter, and thermal saturation of the \cii line emission (e.g., \citealp{2001ApJ...561..766M};  \citealp{2016MNRAS.463.2085M}; \citealp{2017ApJ...846...32D}; \citealp{2018ApJ...861...95H}). Because this so-called \cii deficit is prominent in the central region of galaxies (e.g., \citealp{2017ApJ...834....5S}; \citealp{2019ApJ...870...80L}; \citealp{2020ApJ...904..130V}), it is possible that the larger sizes measured in \cii line emission compared to FIR continuum emission accompany the radial dependence on the \cii deficit.

To inspect this, we stacked ten single galaxies, in which both \cii line and FIR continuum emission is detected. The stacking was performed in visibility using the {\tt CASA/concat} task. The phase centers of individual visibility data are shifted to match the coordinates at which the \cii line emission peaks. We image both stacked \cii line and FIR continuum emission using the {\tt CASA/tclean} task with natural weighting, and clean down to $1.5\sigma$ noise level by applying $3''$ aperture mask, where $\sigma$ is the rms noise level of the dirty image. We show the normalized surface density profiles of the stacked clean image in the left panel of Figure\,\ref{fig:Fig10}. {Both \cii line and FIR continuum emission exhibit an extended structure beyond the PSF Gaussian profile with an effective radius of $1.97\pm0.17$\,kpc and $1.53\pm0.22$\,kpc, respectively. Beyond a beam-convolved radius of $1''$, the surface densities of  \cii line and FIR continuum emission become comparable to the surface densities derived from setting random apertures, and the emission is likely not genuine. The IR surface density profile shows a small bump at $1''$. By inspecting the stacked image, we confirm that this bump results from the elongated structure at the 2-3$\sigma$ significance level of the FIR continuum emission. We therefore do not attribute the bump at $1''$ to mergers or halo structure.} 

In the right panel of Figure\,\ref{fig:Fig10}, we show the absolute surface density profiles (left axis) and their ratio (right axis). The $\Sigma_{\mathrm{[CII]}}/\Sigma_{\mathrm{IR}}$ profile fluctuates around the global value of $8\times10^{-4}$, with a gradual increase until $0\farcs5$ (3\,kpc) and subsequent decrease at least up to $\sim0\farcs8$ (5\,kpc). The global value of $\Sigma_{\mathrm{[CII]}}/\Sigma_{\mathrm{IR}}=8\times10^{-4}$ is in fact smaller than the expected value of $\Sigma_{\mathrm{[CII]}}/\Sigma_{\mathrm{IR}}=(1-4)\times10^{-3}$ (e.g., \citealp{2018ApJ...861...94H}; \citealp{2018MNRAS.481.1976Z}) for the CRISTAL galaxies with $\Sigma_{\mathrm{IR}}\sim10^{10}-10^{11}L_{\odot}/\mathrm{kpc}^{2}$ \citep{2024A&A...690A.197M}. We attribute this lower ratio in the stacked image to the IR-bright galaxies in our sample (CRISTAL-21 and CRISTAL-24), which are an order of magnitude brighter in IR luminosity than the rest of the CRISTAL galaxies. However, our interest here is the radial variation in $\Sigma_{\mathrm{[CII]}}/\Sigma_{\mathrm{IR}}$, and so we have confirmed that the overall trend is unchanged when such exceptionally IR-bright galaxies are excluded from the stacking. Because the radial variations in $\Sigma_{\mathrm{[CII]}}/\Sigma_{\mathrm{IR}}$ are not monotonically decreasing toward the center and instead are relatively flat, the \cii deficit is not particularly strong in the central region. This may be at odds with nearby galaxies \citep{2017ApJ...834....5S}, but consistent with some of the quasar-host galaxies at $z\gtrsim6$ \citep{2020ApJ...904..130V} and the individual CRISTAL galaxies (\citealp{2021A&A...649A..31H}; \citealp{2023A&A...669A..46P}; \citealp{2024ApJ...976...70L}). In summary, we suggest that radial variation in the \cii deficit is not the primary cause of the extended \cii line emission.

\begin{figure*}[ht!]
    \centering
    \includegraphics[width=0.95\linewidth]{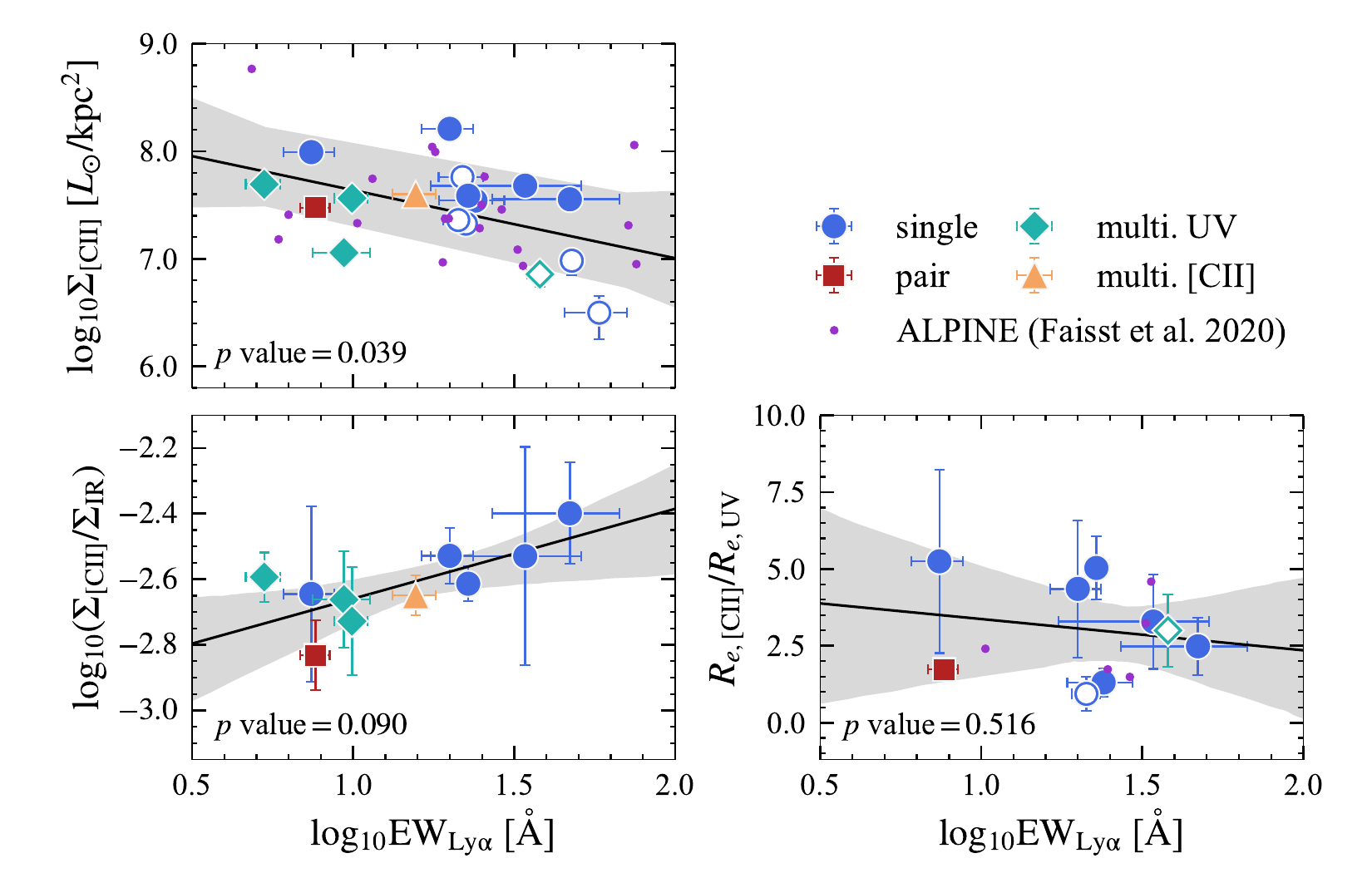}
    \caption{\cii surface density (top left), [C\,{\sc ii}]-to-IR surface density ratio (bottom left), and [C\,{\sc ii}]-to-UV size ratio (bottom right) of the CRISTAL galaxies as a function of the rest-frame EW of Ly$\alpha$ emission. The Ly$\alpha$ properties are taken from \cite{2020ApJS..247...61F}. The galaxies without dust continuum detection are denoted by open symbols. Small purple circles shown in the top left and bottom right panels are the ALPINE galaxies in which \cii surface density, [C\,{\sc ii}]-to-UV size ratio \citep{2020ApJ...900....1F}, and Ly$\alpha$ EW \citep{2020ApJS..247...61F} are available. We show the linear regression fitting (Table\,\ref{table:Table3}) in the solid black line, with the shaded region representing the 95\,\% confidence interval. The $p$\,value of a Spearman's rank correlation test is shown in the lower-left corner of each panel. }
    \label{fig:Fig11}
\end{figure*}

\subsubsection{Shocked gas}
\label{subsubsec:Section4.3.4}

The alternative mechanism that can enhance the \cii line emission without star formation activity is shock heating produced in a merging system or in a galaxy group (\citealp{2013ApJ...777...66A}; \citealp{2018ApJ...855..141P}; \citealp{2023ApJ...957...83F}): kinetic energy from shocks and turbulence is injected into the gas and excite the ionized carbon by collisions. We refer the readers to \cite{2024arXiv240303379P} for detailed discussion on the argument that the \cii line emission in CRISTAL-05 is likely to be enhanced by shock heating. In short, the enhanced \cii line emission due to shock heating is suggested from evidence of merging activity and a high 
[C\,{\sc ii}]-to-IR luminosity ratio ($L_{\mathrm{[CII]}}/L_{\mathrm{IR}}\gtrsim10^{-2}$). The latter can also be tested by comparing the \cii surface density to SFR surface density. 
 
Among the CRISTAL galaxies except for CRISTAL-05 \citep{2024arXiv240303379P}, the pair and multiple-UV systems are the likely candidates that may emit \cii line emission induced by shock heating. However, as we show in Table\,\ref{table:Table2} and discuss at the beginning of Section\,\ref{subsec:Section4.3}, most of these systems have a moderate size ratio between the \cii line and FIR continuum ($R_{e,{\mathrm{[CII]}}}/R_{e,{\mathrm{FIR}}}\sim0.7-1.5$), so their \cii line emission can mostly be explained by PDRs. Two exceptions are CRISTAL-04a ($R_{e,{\mathrm{[CII]}}}/R_{e,{\mathrm{FIR}}}=2.1\pm1.2$) and CRISTAL-13 ($R_{e,{\mathrm{[CII]}}}/R_{e,{\mathrm{FIR}}}=2.7\pm2.1$), but neither of them shows high [C\,{\sc ii}]-to-IR luminosity ratio ($L_{\mathrm{[CII]}}/L_{\mathrm{IR}}\simeq 3\times10^{-3}$). Shocked gas heating is not likely the driver of extended \cii line emission.

\subsection{Connection to Ly$\alpha$ and H$\alpha$  emission}
\label{subsec:Section4.4}

\cite{2018ApJ...859...84H} find an anti-correlation between \cii luminosity and Lyman-$\alpha$ line (Ly$\alpha$ line, $\lambda=1216$\,$\mathrm{\AA}$) equivalent width (EW) for Ly$\alpha$ emitters at $z=4-7$. This anti-correlation is likely to exist for each galaxy population, such as Lyman-$\alpha$ emitters and star-forming main sequence galaxies, since typical \cii luminosity at fixed Ly$\alpha$ EW differs across galaxy populations \citep{2021A&A...654A..37D}. However, \cite{2022MNRAS.517.5642E} do not find such anti-correlation for UV-bright galaxies at $z\simeq7$, possibly due to either a limited dynamic range of Ly$\alpha$ EW ($<20$\AA) or insufficient statistics. 

\begin{table}[t]
\renewcommand{\arraystretch}{1.4}
\caption{Summary of the best-fit parameters in Figure\,\ref{fig:Fig11}.}
\label{table:Table3}
\centering
  \begin{tabular}{ccc}
    \hline\hline 
    $y$ & $m$ & $b$ \\
    \hline
    $\log_{10}{\Sigma_{\mathrm{[CII]}}}$ & $-0.633$ & $8.27$ \\
    $\log_{10}({\Sigma_{\mathrm{[CII]}}/\Sigma_{\mathrm{IR}}})$ & $0.274$ & $-2.93$ \\
    $R_{e,\mathrm{[CII]}}/R_{e,\mathrm{UV}}$ & $-1.02$ & $4.39$ \\
    \hline
  \end{tabular}
  \tablefoot{{Slope ($m$) and intercept ($b$) of the linear regression fittings performed in Figure\,\ref{fig:Fig11}. Parameters are determined from the form of $y=m\log_{10}{\mathrm{EW_{Ly\alpha}}}+b$.}}
\end{table}

To study how the \cii line properties of our sample correlate with Ly$\alpha$ emission, we plot the \cii line surface density ($\Sigma_{\mathrm{[CII}]}=L_{\mathrm{[CII}]}/2\pi R_{e,{\mathrm{[CII]}}}^{2}$), the [C\,{\sc ii}]-to-IR surface density ratio ($\Sigma_{\mathrm{[CII}]}/\Sigma_{\mathrm{IR}}$) within the \cii effective radius, and the [C\,{\sc ii}]-to-UV size ratio ($R_{e,\mathrm{[CII]}}/R_{e,\mathrm{UV}}$) as a function of the Ly$\alpha$ EW ($\mathrm{EW_{Ly\alpha}}$) in Figure\,\ref{fig:Fig11}. We estimate the IR luminosity within the \cii radius based on the best-fit model in the visibility modeling. The values of $\mathrm{EW_{Ly\alpha}}$ are presented in the catalogs of both \cite{2020ApJS..247...61F} and \cite{2020A&A...643A...6C}. Since \cite{2020A&A...643A...6C} report the $\mathrm{EW_{Ly\alpha}}$ value only for the galaxies with \cii line detection in the ALPINE Survey, we adopt the values from \cite{2020ApJS..247...61F} to increase the sample size. Here, we convert the $\mathrm{EW_{Ly\alpha}}$ reported in the observed-frame to the rest-frame by applying a correction factor of $(1+z)^{-1}$. In agreement with previous findings, we find a decreasing trend of the \cii line surface density with increasing $\mathrm{EW_{Ly\alpha}}$, with a Spearman's rank correlation coefficient of $-0.51$ and a $p$\,value of $0.039$. Moreover, we find the [C\,{\sc ii}]-to-IR surface density ratio positively correlates with $\mathrm{EW_{Ly\alpha}}$ with a correlation coefficient of $0.56$ and a $p$\,value of $0.090$. The size ratio of \cii line and UV radii appears to decrease with $\mathrm{EW_{Ly\alpha}}$, which aligns with the results from \cite{2020ApJ...900....1F}, although a Spearman's rank correlation test give a $p$\,value of 0.516. {We perform a linear regression fitting for each correlation and present the fitting results in Table\,\ref{table:Table3}.} As a side note, we do not find a correlation between \cii line radius and $\mathrm{EW_{Ly\alpha}}$.

The correlations we present in Figure\,\ref{fig:Fig11} imply that there is physical interplay between \cii line, dust, and Ly$\alpha$ emission. 
Noticeably, we find that three galaxies (CRISTAL-01a, 12, and 17) with the lowest \cii surface density with $\Sigma_{\mathrm{[CII]}}<10^{7} L_{\odot}/\mathrm{kpc^{2}}$ have large Ly$\alpha$ EW ($\mathrm{EW_{Ly\alpha}}\gtrsim10^{1.5}$\,\AA) and are not detected in FIR continuum emission \citep{2024A&A...690A.197M}. On the other hand, galaxies with small Ly$\alpha$ EW ($\mathrm{EW_{Ly\alpha}}\lesssim10^{1.5}$\,\AA) exhibit a high \cii surface density ($\Sigma_{\mathrm{[CII]}}>10^{7} L_{\odot}/\mathrm{kpc^{2}}$) and a low [C\,{\sc ii}]-to-IR surface density ratio ($\Sigma_{\mathrm{[CII}]}/\Sigma_{\mathrm{IR}}\lesssim10^{-2.5}$); in other words, a \cii deficit. These ratios indicate that the low dust content in the ISM of these galaxies facilitates the escape of Ly$\alpha$ photons and vice versa. In such ISM conditions, the ionization parameter is expected to be higher compared to the galaxies with strong FIR continuum detection. Simultaneously, the limited amount of dust likely leads to low \cii surface density by reducing the efficiency of photoelectric heating.

It is known that the velocity offsets between the systemic velocity (measured from \cii line or optical emission lines) and the velocity of the peak of Ly$\alpha$ line anti-correlate with $\mathrm{EW_{Ly\alpha}}$ (\citealp{2013ApJ...765...70H,2019PASJ...71...71H}; \citealp{2014ApJ...795...33E}; \citealp{2022ApJ...929L...9V}), which has also been confirmed for the ALPINE sample \citep{2020A&A...643A...6C}. As large velocity offsets indicate high H\,{\sc i} column density, the galaxies with small $\mathrm{EW_{Ly\alpha}}$ have higher H\,{\sc i} column densities than the ones with large $\mathrm{EW_{Ly\alpha}}$, which may explain extended \cii line emission if it arises from the atomic gas (Section\,\ref{subsec:Section4.3}). Therefore, although it is not conclusive whether the weak anti-correlation between $R_{e,\mathrm{[CII]}}/R_{e,\mathrm{UV}}$ and Ly$\alpha$ EW is real, the large Ly$\alpha$ EW, low H\,{\sc i} column density, small amount of dust, and small $R_{e,\mathrm{[CII]}}/R_{e,\mathrm{UV}}$ are consistent with each other, and larger fraction of the \cii line emission may arise from the ionized gas in such galaxies (Section\,\ref{subsubsec:Section4.3.2}).

However, we note that the Ly$\alpha$ properties currently available are taken by slit spectroscopy \citep{2020A&A...643A...6C}, and only global properties without spatial information can be inferred. We thus expect that future Ly$\alpha$ observations by integral field spectroscopy (IFS) would be crucial to compare the spatial extent and morphology between \cii line and Ly$\alpha$ emission. 

Another fundamental tracer emitted from hydrogen atoms is the Balmer-$\alpha$ line (H$\alpha$ line, $\lambda=6563$\AA). H$\alpha$ line traces an instantaneous star formation activity ($\sim10-100$\,Myr), so if the H$\alpha$ line is more extended than continuum emission tracing star-formation activity over $\sim100$\,Myr, this would support the idea that these galaxies are in a phase of ``inside-out'' growth  and neutral gas at larger radii is heated by recent star formation activity. \cite{2015ApJ...800....1H} show that \cii line emission can arise and hold the [C\,{\sc ii}]-SFR relation with a short duration (2\,Myr) of star formation activity with high heating efficiency from FUV to \cii line emission ($\epsilon_{\mathrm{h}}\simeq1-3$\,\%), suggesting the possibility that recent star formation activity causing the extended \cii line emission may be traced by H$\alpha$ line emission.

As JWST has recently enabled us to study the spatial distribution of the H$\alpha$ line at comparable redshifts to our sample (e.g., \citealp{2023ApJ...958...64B}; \citealp{2024MNRAS.533.4287U}; \citealp{2024A&A...690A..64M}; \citealp{2024ApJ...976L..27N}), it is of great interest to compare the morphological and kinematical properties of H$\alpha$ and \cii lines \citep{2024A&A...685A..72K}. {Two of the galaxies among our sample (CRISTAL-20 and CRISTAL-22) have been observed in JWST/NIRSpec IFU mode (\citealp{2024arXiv240719008P}; \citealp{2024arXiv240512955J}), offering new insights into the ionized gas properties of typical star-forming galaxies at high redshift. In CRISTAL-20, \cite{2024arXiv240719008P} reveal ionized outflows traced by the [O\,{\sc iii}]5007\AA\,\,and H$\alpha$ lines, which are co-spatial to the neutral outflows traced by \cii line. \cite{2024arXiv240512955J} discover brighter H$\alpha$ line emission in CRISTAL-22b than CRISTAL-22a, whereas dust continuum is weaker, suggesting different evolutionary phases among the complex merging system. A future systematic census of ionized gas properties in the CRISTAL galaxies with JWST will provide unprecedented insights into the spatial correlation between neutral and ionized gas.}

\section{Conclusions} \label{sec:Section5}
We have presented the \cii size measurements and size comparisons with other tracers for typical SFGs at $z=4-6$ taken by the CRISTAL Survey. The data of the survey benefits from the deep sensitivity ($1\sigma\sim 0.1-0.2$ mJy per 20~km/s channel bin) to search faint extended emission and high spatial resolution ($\theta_{\rm beam}\sim0\farcs3$), which enables us to identify and characterize multiple galaxies in proximity. Our main findings are as follows:

\begin{enumerate}
\item \textit{\cii sizes are similar among various merger states.} We performed size measurements of \cii line emission via visibility modeling analysis. The structural parameters of 33 components are reported in Table\,\ref{table:Table1}, including pairs (e.g., CRISTAL-07ab) and complex UV systems (e.g., CRISTAL-02, CRISTAL-10). In spite of a variety of merger states, we do not statistically confirm the difference in \cii sizes between single and pair ($p$\,value = 0.23)/multiple-UV ($p$\,value = 0.12) systems. \\

\item \textit{Extended \cii line emission revealed by size comparison.} Based on the comparisons between \cii line radius and rest-frame UV, and FIR radii, we find average size ratios and standard deviations of $\langle R_{e,{\rm [CII]}}/R_{e,{\rm UV}}\rangle=2.90\pm1.40$ and $\langle R_{e,{\rm [CII]}}/R_{e,{\rm FIR}}\rangle=1.54\pm0.63$, respectively. This indicates that \cii line emission is approximately twice more extended than the main star-forming regions traced by rest-frame UV and FIR continuum emission. \\

\item \textit{\cii line emission as a single extended disk component.} We assessed whether the \cii line emission can be characterized by a disk with a secondary halo structure {by performing both single and double component fittings in visibility. While we have obtained the two-component model ($R_{e,\mathrm{gal}}=1.91\pm0.20$\,kpc, $R_{e,\mathrm{halo}}=9.17\pm28.2$\,kpc), the flux density of ``\cii halo'' is weaker than the typical uncertainty of the visibility data in short $uv$ distances ($\sim20$\,k$\lambda$), and the single extended  disk component ($R_{e}=1.97\pm0.10$\,kpc) characterizes the observed \cii line emission with an equal level of statistical support.} In the stacked residual visibility, we did not detect any significant flux excess, thus we conclude that the \cii line emission in our sample can be adequately characterized by a single disk component. \\

\item \textit{Origin of \cii line emission.} While the \cii line emission in about half of our sample can be explained by PDRs associated with the star formation activity, the extent of \cii line emission in the rest of the sample appears to require an additional source of excitation. The most likely is \cii line emission coming from the diffuse medium, either neutral atomic gas or ionized gas, although the effect of such components at high redshift is unsettled from a theoretical perspective. We argue that the diversity in \cii sizes may be further caused by different merger  {history and the existence of extraplanar gas and dust}. In contrast, the \cii deficit in the central region of the galaxies, and shocked gas heating induced by mergers are not likely to be responsible for the extended \cii line emission. \\

\item \textit{Correlations with the Ly$\alpha$ properties.} {We find a negative correlation between \cii surface density ($\Sigma_{\mathrm{[CII]}}$) and Ly$\alpha$ EW ($p$\,value $=0.039$), and a tentative positive correlation between [C\,{\sc ii}]-to-IR surface density ratio ($\Sigma_{\mathrm{[CII]}}/\Sigma_{\mathrm{IR}}$) and Ly$\alpha$ EW ($p$\,value $=0.090$).} These correlations between \cii and Ly$\alpha$ properties can be understood by the amount of gas and dust content, which is essential for producing the \cii line emission via photoelectric heating. {High \cii surface density suggests high neutral gas column density, leading to strong attenuation of Ly$\alpha$ emission (low Ly$\alpha$ EW).} We also find a possible negative correlation between [C\,{\sc ii}]-to-UV size ratio ($R_{e,\mathrm{[CII]}}/R_{e,\mathrm{UV}}$) and Ly$\alpha$ EW. If the correlation is real, this would support the scenario that extended \cii line emission at large radii is arising from the atomic gas component. Future IFS studies of both Ly$\alpha$ and H$\alpha$ lines are expected to explain the variations in the \cii line and dust properties among the typical SFGs at this epoch. \\
\end{enumerate}

\begin{acknowledgements} \\
      {We thank the anonymous referee for valuable comments that improved the content of this paper.}
      R.I. would like to thank Seiji Fujimoto, Mahsa Kohandel, Rebecca C. Levy, Zongnan Li, Yuichi Matsuda, Marcel Neeleman, and Andrea Pallottini for illuminating discussions and suggestions.
      R.I. is supported by Grants-in-Aid for Japan Society for the Promotion of Science (JSPS) Fellows (KAKENHI Grant Number 23KJ1006). 
      K.-i.T. is supported by JSPS KAKENHI grant No. 23K03466. 
      M.A. acknowledges support from the ANID BASAL project FB210003. 
      I.D.L. has received funding from the European Research Council (ERC) under the European Union's Horizon 2020 research and innovation programme DustOrigin (ERC-2019-StG-851622) and from the Flemish Fund for Scientific Research (FWO-Vlaanderen) through the research projects G023821N and G0A1523N. 
      R.H.-C. thanks the Max Planck Society for support under the Partner Group project "The Baryon Cycle in Galaxies" between the Max Planck for Extraterrestrial Physics and the Universidad de Concepci\'{o}n. R.H-C. also gratefully acknowledge financial support from ANID BASAL projects FB210003. 
      R.A.A.B. acknowledges support from an STFC Ernest Rutherford Fellowship [grant number ST/T003596/1]. 
      E.d.C acknowledges support from the Australian Research Council (projects DP240100589 and CE170100013). 
      R.L.D. is supported by the Australian Research Council through the Discovery Early Career Researcher Award (DECRA) Fellowship DE240100136 funded by the Australian Government. 
      T.D.S. was supported by the Hellenic Foundation for Research and Innovation (HFRI) under the ``2nd Call for HFRI Research Projects to support Faculty Members \& Researchers'' (Project Number: 3382). 
      A.F. acknowledges support from the ERC Advanced Grant INTERSTELLAR H2020/740120. 
      M.K. was supported by the ANID BASAL project FB210003. 
      M.S. was financially supported by Becas-ANID scholarship \#21221511, and also acknowledges ANID BASAL project FB210003. 
      K.T. was supported by ALMA ANID grant number 31220026 and by the ANID BASAL project FB210003. 
      H.\"{U}. gratefully acknowledges support by the Isaac Newton Trust and by the Kavli Foundation through a Newton-Kavli Junior Fellowship. 
      V.V. acknowledges support from the ALMA-ANID Postdoctoral Fellowship under the award ASTRO21-0062. 
      This paper makes use of the following ALMA data: ADS/JAO.ALMA\#2017.1.00428.L, \#2018.1.01359.S, \#2018.1.01605.S, \#2019.1.00226.S, \#2019.1.01075.S,
      and \#2021.1.00280.L. ALMA is a partnership of ESO (represent- ing its member states), NSF (USA), and NINS (Japan), together with NRC (Canada), MOST and ASIAA (Taiwan), and KASI (Republic of Korea), in cooperation with the Republic of Chile. The Joint ALMA Observatory is operated by ESO, AUI/NRAO, and NAOJ. We thank the ALMA staff and, in particular, the EA-ARC staff for their support. Data analyses were carried out in part on the Multiwavelength Data Analysis System operated by the Astronomy Data Center (ADC) at the National Astronomical Observatory of Japan.
\end{acknowledgements}

\bibliographystyle{aa.bst}
\bibliography{bibtex_ikeda} 

\begin{appendix}
\section{The sample name and postage stamps of the fit images}
\label{sec:AppendixA}

\begin{table}[ht]
\renewcommand{\arraystretch}{1.4}
{\caption{Other names and references of the CRISTAL galaxies.} \label{table:TableA1}}
\centering
  \begin{tabular}{lll}
    \hline\hline
    ID & Other names & Ref. \\
    \hline
CRISTAL-01a & DC842313 & 1 \\
CRISTAL-01b	& -- & -- \\
CRISTAL-02	& DC848185, HZ6, LBG-1 & 1, 2, 3 \\
CRISTAL-03	& DC536534, HZ1 & 1, 2 \\
CRISTAL-04a	& vc5100822662, DC514583 & 1 \\ 
CRISTAL-04b	& vc5100822662, DC514583 & 1 \\
CRISTAL-05	& DC683613, HZ3 & 1, 2 \\
CRISTAL-06a	& vc5100541407 & 1 \\
CRISTAL-06b	& vc5100541407 & 1 \\
CRISTAL-07a	& DC873321, HZ8 & 1,2 \\
CRISTAL-07b	& DC873321, HZ8W & 1,2 \\
CRISTAL-07c	& -- & -- \\
CRISTAL-08	& ve530029038, CG15 & 1 \\
CRISTAL-09	& DC519281 & 1 \\
CRISTAL-10	& DC417567, HZ2 & 1, 2 \\
CRISTAL-11	& DC630594 & 1 \\
CRISTAL-12	& CG21 & 1 \\
CRISTAL-13	& vc5100994794 & 1 \\
CRISTAL-14	& DC709575 & 1 \\
CRISTAL-15	& vc5101244930 & 1 \\
CRISTAL-16	& CG38 & 1 \\
CRISTAL-17	& DC742174 & 1 \\
CRISTAL-18	& vc5101288969, DC679410 & 1 \\
CRISTAL-19	& DC494763 & 1 \\
CRISTAL-20	& DC494057, HZ4 & 1, 2 \\
CRISTAL-21	& HZ7 & 1 \\
CRISTAL-22a	& HZ10, HZ10-C & 2, 4 \\
CRISTAL-22b	& HZ10, HZ10-W & 2, 4 \\
CRISTAL-23a	& DC818760 & 1 \\
CRISTAL-23b	& DC818760 & 1 \\
CRISTAL-23c & -- & -- \\
CRISTAL-24	& DC873756 & 1 \\
CRISTAL-25	& vc5101218326 & 1 \\
\hline
  \end{tabular}
\tablebib{
1. \cite{2020A&A...643A...2B}; 2. \cite{2015Natur.522..455C}; 3. \cite{2014ApJ...796...84R}; 4. \cite{2024A&A...691A.133V}.}
\end{table}

\begin{figure*}[h!]
    \centering
	\includegraphics[width=\linewidth]{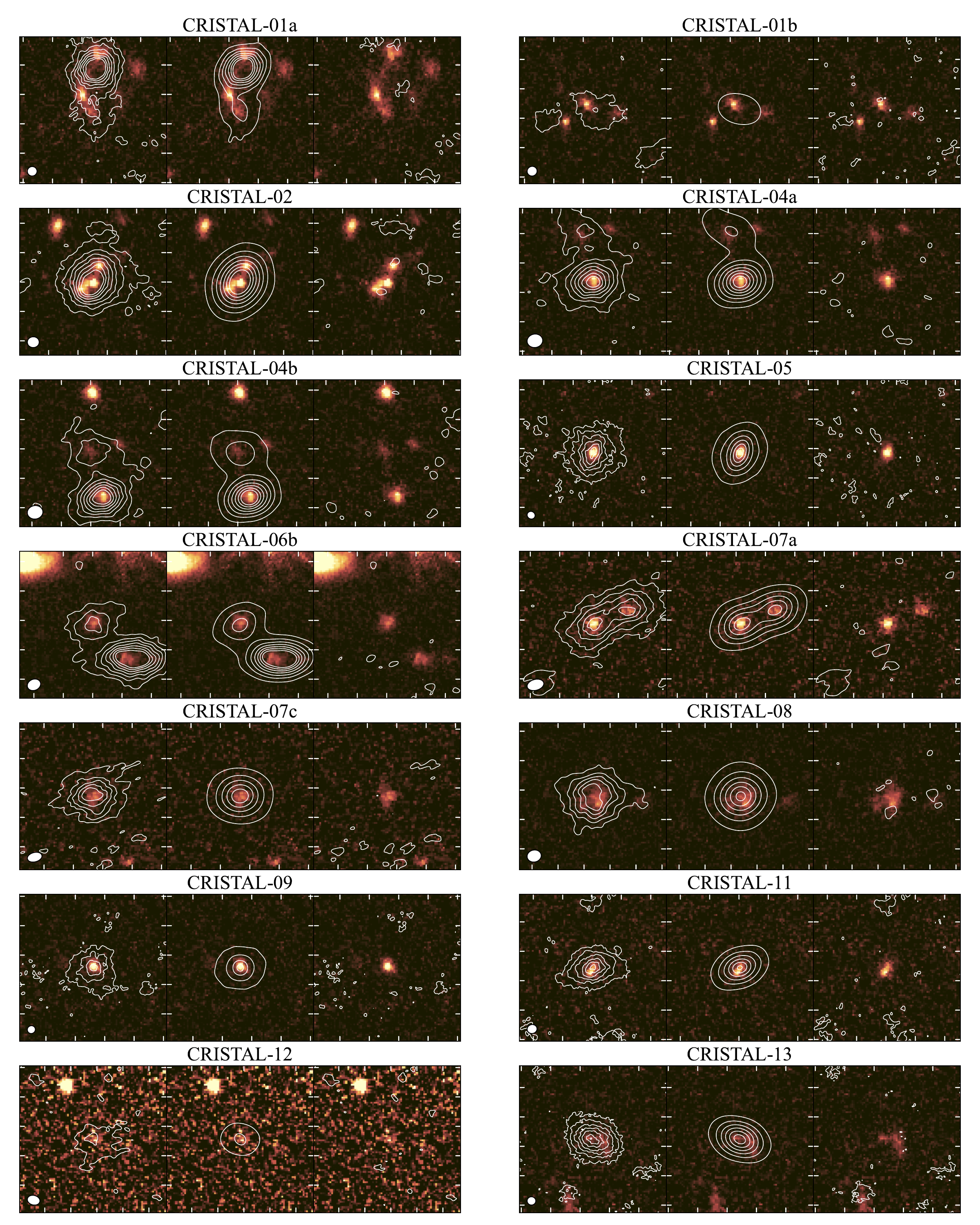}
    \caption{Visualization of the visibility modeling of the \cii line emission in images. The \cii line contours are overlaid on the HST/WFC3 F160W image. The left panel shows the dirty image of a single cube used for the visibility modeling (Section~\ref{subsec:Section3.2}). {The FWHM of the dirty beam is shown as a white ellipse in the lower-left corner.} The middle and right panel show the best-fit model and the residual created by subtracting the model from the data, respectively. A $5''\times5''$ region is shown. The contours start at $2\sigma$ and increase in steps of $3\sigma$ until $20\sigma$.}
    \label{fig:FigA1}
\end{figure*}

\begin{figure*}[h!]
    \includegraphics[width=1.005\linewidth]{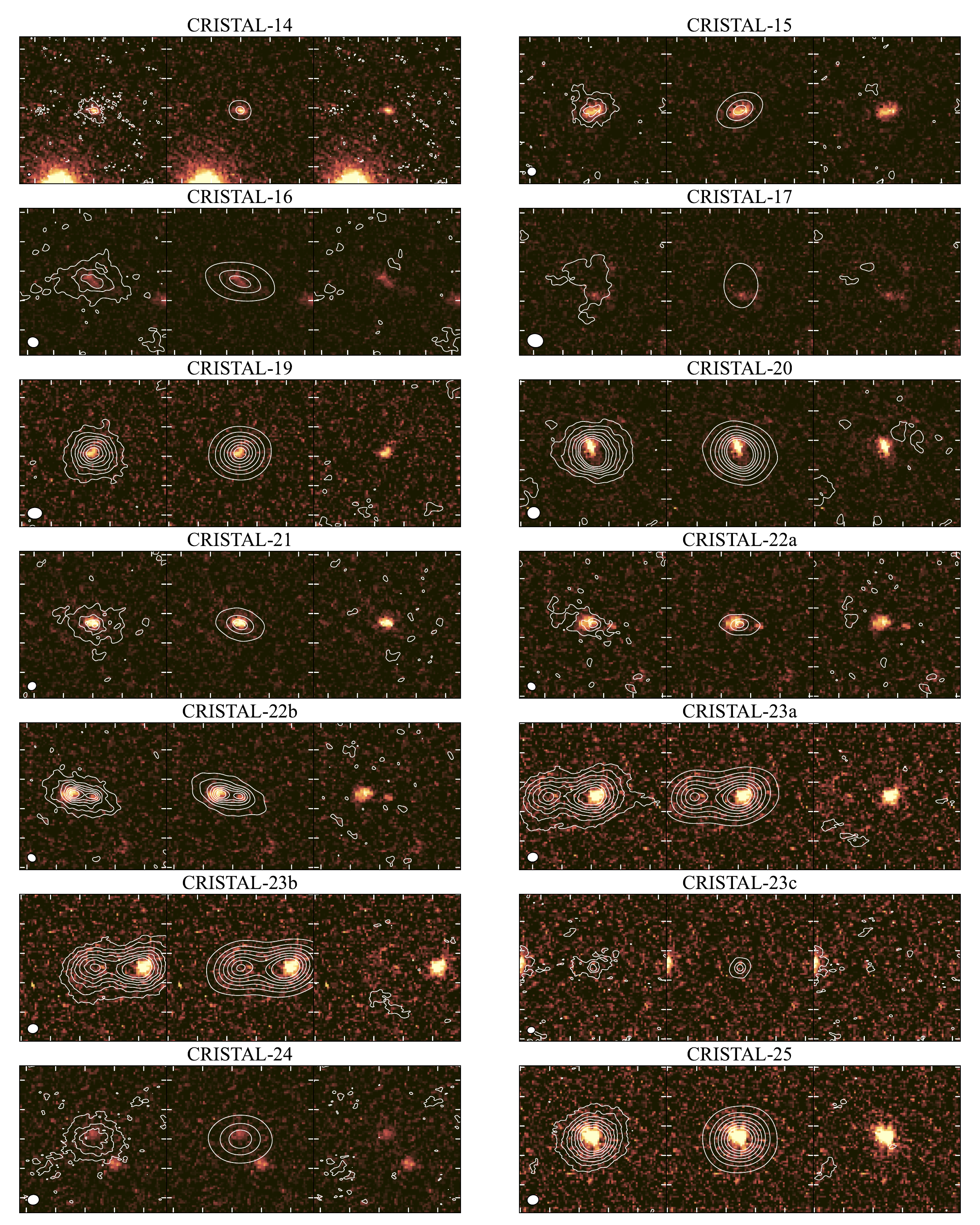}
    \caption{continued.}
    \label{fig:FigA2}
\end{figure*}

\section{Visibility modeling with the ALPINE data}
\label{sec:AppendixC}
In this section, we briefly discuss how our visibility modeling improved by adding the data taken from the CRISTAL survey. In Table\,\ref{table:TableC1}, we show the results of visibility modeling using the ALPINE data alone. We followed the same procedure described in Section~\ref{subsec:Section3.2}. Out of the 33 \cii components reported in this work, the ALPINE data was used for 24 components, and we have successfully modeled 21 components. The fittings did not converge for the rest of the three components, either due to the compactness or faintness of the \cii line emission. Some of the merging systems (e.g., CRISTAL-04a and CRISTAL-04b pair) were able to be modeled as two distinct galaxies if we allow a circular exponential model. 

Figure\,\ref{fig:FigC1} compares the fit values between two different datasets. {A slope and $p$\,value based on a linear regression fitting are shown. For the linear regression fitting of the minor-to-major axis ratios, we exclude the galaxies for which the minor-to-major axis ratio is unconstrained by the ALPINE data ($q_{\mathrm{ALPINE}}=1$). We find that all three correlations (total flux, effective radius, and minor-to-major axis ratio) show a good agreement with the $p$\,value smaller than 0.05, demonstrating that the ALPINE data alone could constrain three parameters well. However, the uncertainties of flux and radius become on average 2.1 and 1.8 times smaller, respectively, when the CRISTAL data is included.} One major improvement is that we have successfully constrained the minor-to-major axis ratio for more than 90\,\% of the sample, while less than half cannot be modeled when only low-resolution data is used. This could be either due to the difference in the sensitivities or the spatial resolutions of the data.

\begin{figure*}[ht!]
    \center
    \includegraphics[width=0.95\linewidth]{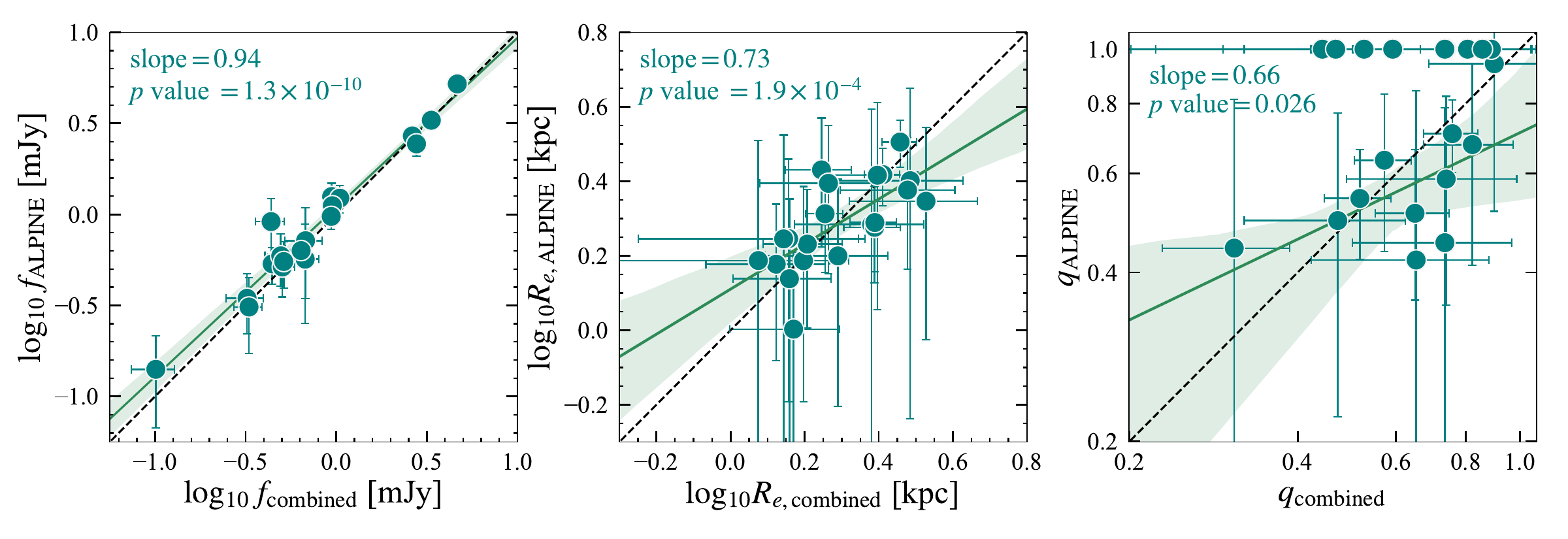}
    {\caption{Comparison of total flux (left), effective radius (middle), and axis ratio (right) measured from the visibility modeling using different datasets. A slope and $p$\,value based on a linear regression fitting are shown in the top-left corner of each panel. We exclude the galaxies with $q_{\mathrm{ALPINE}}=1$ for the fitting of minor-to-major axis ratio.} \label{fig:FigC1}}
\end{figure*}

\begin{table*}[h!]
\renewcommand{\arraystretch}{1.4}
\caption{Structural parameters of the \cii line emission measured from the ALPINE data alone.}
\label{table:TableC1}
\centering
  \begin{tabular}{lcccc}
    \hline\hline
    ID & $I_{\rm [CII]}$ & $R_{e, { \rm [CII]}}$ & $q_{\rm [CII]}$ & PA$_{\rm [CII]}$ \\
     & (Jy~km/s) & (kpc) & & (deg) \\
    \hline
CRISTAL-01a & $0.72\pm0.37$ & $2.52\pm1.94$ & -- & -- \\
CRISTAL-01b & $0.57\pm0.32$ & $1.92\pm2.00$ & -- & -- \\
CRISTAL-02 & $2.70\pm0.24$ & $2.62\pm0.46$ & $0.54\pm0.12$ & $156\pm9$ \\
CRISTAL-03 & $0.54\pm0.12$ & $1.50\pm0.68$ & -- & -- \\
CRISTAL-04a & $1.25\pm0.23$ & $2.70\pm1.02$ & $0.94\pm0.43$ & $51\pm226$ \\
CRISTAL-04b & $0.35\pm0.12$ & $2.22\pm1.28$ & -- & -- \\
CRISTAL-05 & $1.01\pm0.18$ & $1.54\pm0.89$ & $0.44\pm0.37$ & $146\pm25$ \\
CRISTAL-07a & $0.59\pm0.19$ & $2.48\pm1.06$ & -- & -- \\
CRISTAL-07b & $0.51\pm0.16$ & $1.58\pm0.96$ & -- & -- \\
CRISTAL-09a & -- & -- & -- & -- \\
CRISTAL-10 & $0.55\pm0.16$ & $2.61\pm1.47$ & $0.45\pm0.34$ & $163\pm23$ \\
CRISTAL-11 & $1.23\pm0.21$ & $1.38\pm0.88$ & $0.42\pm0.42$ & $83\pm24$ \\
CRISTAL-12 & $0.14\pm0.07$ & $1.54\pm1.69$ & -- & -- \\
CRISTAL-13 & $1.12\pm0.17$ & $1.70\pm0.69$ & $0.49\pm0.27$ & $63\pm19$ \\
CRISTAL-14 & -- & -- & -- & -- \\
CRISTAL-15 & $0.91\pm0.31$ & $1.76\pm1.12$ & -- & -- \\
CRISTAL-16 & $0.31\pm0.14$ & $1.76\pm1.59$ & -- & -- \\
CRISTAL-17 & -- & -- & -- & -- \\
CRISTAL-19 & $0.63\pm0.11$ & $1.01\pm0.61$ & -- & -- \\
CRISTAL-20 & $0.97\pm0.10$ & $2.06\pm0.63$ & $0.63\pm0.20$ & $37\pm18$ \\
CRISTAL-23a & $5.21\pm0.42$ & $3.20\pm0.47$ & $0.71\pm0.10$ & $115\pm13$ \\
CRISTAL-23b & $2.44\pm0.36$ & $1.89\pm0.55$ & $0.68\pm0.26$ & $149\pm27$ \\
CRISTAL-24 & $13.62\pm2.08$ & $2.38\pm0.92$ & $0.59\pm0.24$ & $108\pm20$ \\
CRISTAL-25 & $3.29\pm0.23$ & $1.95\pm0.51$ & $0.51\pm0.15$ & $33\pm10$ \\
\hline
  \end{tabular}
\end{table*}

\end{appendix}

\end{document}